\documentclass[abstract=true,headings=standardclasses]{scrartcl}

\usepackage[utf8]{inputenc}
\usepackage{helvet,fancybox}
\usepackage{subcaption}
\usepackage{amsfonts}
\usepackage{graphicx}
\usepackage{longtable}
\usepackage{pdfpages}
\usepackage[english]{babel}
\usepackage{url}
\usepackage{a4wide}
\usepackage{dsfont}
\usepackage{calrsfs}
\usepackage{csvsimple}
\usepackage{siunitx}
\usepackage[acronyms]{glossaries}
\usepackage{booktabs}

\DeclareFontFamily{OT1}{pzc}{}
\DeclareFontShape{OT1}{pzc}{m}{it}{<-> s * [1.10] pzcmi7t}{}
\DeclareMathAlphabet{\mathpzc}{OT1}{pzc}{m}{it}

\newcommand{\G}{\mathbf{G}}
\newcommand{\FIT}{\mathrm{FIT}}

\DeclareMathAlphabet{\pazocal}{OMS}{zplm}{m}{n}

\DeclareMathOperator{\dddiv}{\boldsymbol{div}}

\newcommand{\Q}{\mathbf{Q}}

\newtheorem{rmk}{Remark}

\usepackage{hyperref}
\usepackage{bbm}
\usepackage{xcolor}
\usepackage{amsmath}
\usepackage{xspace}
\usepackage{upgreek}
\allowdisplaybreaks

\newacronym{DAE}{\protect DAE}{differential algebraic equation}
\newacronym{FE}{\protect FE}{finite element}
\newacronym{FEM}{\protect FEM}{finite element method}
\newacronym{FIT}{\protect FIT}{finite integration technique}
\newacronym{PEC}{\protect PEC}{perfect electric conducting}

\catcode`@=11
\newcommand{\frownfill}{\ensuremath{\scriptscriptstyle\m@th\mathord\frown}}
\newcommand{\bow}[1]{\ensuremath{\vbox{\m@th\ialign{##\crcr
      \hfil\frownfill\hfil\crcr\noalign{\kern-0.2\p@\nointerlineskip}
      $\hfil\displaystyle{#1}\hfil$\crcr}}}}
\newcommand{\bbow}[1]{\ensuremath{\vbox{\m@th\ialign{##\crcr
     \hfil\frownfill\hfil\crcr\noalign{\kern-0.7\p@\nointerlineskip}
     \hfil\frownfill\hfil\crcr\noalign{\kern-0.3\p@\nointerlineskip}
      $\hfil\displaystyle{#1}\hfil$\crcr}}}}
\newcommand{\bbbow}[1]{\ensuremath{\vbox{\m@th\ialign{##\crcr
     \hfil\frownfill\hfil\crcr\noalign{\kern-0.7\p@\nointerlineskip}
     \hfil\frownfill\hfil\crcr\noalign{\kern-0.7\p@\nointerlineskip}
     \hfil\frownfill\hfil\crcr\noalign{\kern-0.3\p@\nointerlineskip}
      $\hfil\displaystyle{#1}\hfil$\crcr}}}}
\newcommand{\widefrownfill}{\ensuremath{\m@th\mathord\frown}}
\newcommand{\widebow}[1]{\ensuremath{\vbox{\m@th\ialign{##\crcr
      \hfil\widefrownfill\hfil\crcr\noalign{\kern-0.9\p@\nointerlineskip}
      $\hfil\displaystyle{#1}\hfil$\crcr}}}}
\newcommand{\widebbow}[1]{\ensuremath{\vbox{\m@th\ialign{##\crcr
     \hfil\widefrownfill\hfil\crcr\noalign{\kern-1.8\p@\nointerlineskip}
     \hfil\widefrownfill\hfil\crcr\noalign{\kern-0.9\p@\nointerlineskip}
      $\hfil\displaystyle{#1}\hfil$\crcr}}}}
\newcommand{\widebbbow}[1]{\ensuremath{\vbox{\m@th\ialign{##\crcr
     \hfil\widefrownfill\hfil\crcr\noalign{\kern-1.8\p@\nointerlineskip}
     \hfil\widefrownfill\hfil\crcr\noalign{\kern-1.8\p@\nointerlineskip}
     \hfil\widefrownfill\hfil\crcr\noalign{\kern-0.9\p@\nointerlineskip}
      $\hfil\displaystyle{#1}\hfil$\crcr}}}}

\usepackage{authblk}

\setkomafont{title}{\normalfont}
\title{\LARGE Coupled Simulation of Transient Heat Flow and Electric Currents in Thin Wires: Application to Bond Wires in Microelectronic Chip Packaging}
\author[1,2]{Thorben~Casper}
\author[3]{Ulrich~R\"omer}
\author[1,2]{Sebastian~Sch\"ops}
\author[1,2]{Herbert~De~Gersem}
\affil[1]{Institut f\"ur Theorie Elektromagnetischer Felder, Technische Universit\"at Darmstadt, Schlo{\ss}gartenstr. 8, 64289 Darmstadt Germany}
\affil[2]{Graduate School of Computational Engineering, Technische Universit\"at Darmstadt, Dolivostr. 15, 64293 Darmstadt, Germany}
\affil[3]{Institute of Dynamics and Vibrations, Technische Universit\"at Braunschweig, Schleinitzstr. 20, 38106 Braunschweig, Germany}
\date{\vspace{-5ex}}

\newenvironment{keywords}{\begin{trivlist}\item[]{\bfseries Keywords:}}{\end{trivlist}}

\begin{document}

\maketitle

\begin{abstract}
This work addresses the simulation of heat flow and electric currents in thin wires.
An important application is the use of bond wires in microelectronic chip packaging.
The heat distribution is modeled by an electrothermal coupled problem, which poses numerical challenges due to the presence of different geometric scales.
The necessity of very fine grids is relaxed by solving and embedding a 1D sub-problem along the wire into the surrounding 3D geometry.
The arising singularities are described using de Rham\xspace currents.
It is shown that the problem is related to fluid flow in porous 3D media with 1D fractures [C. D'Angelo, SIAM Journal on Numerical Analysis 50.1, pp. 194-215, 2012].
A careful formulation of the 1D-3D coupling condition is essential to obtain a stable scheme that yields a physical solution.
Elliptic model problems are used to investigate the numerical errors and the corresponding convergence rates.
Additionally, the transient electrothermal simulation of a simplified microelectronic chip package as used in industrial applications is presented.
\end{abstract}


\section{Introduction}
\label{sec:introduction}

\begin{figure}[t!]
	\centering 
    \includegraphics{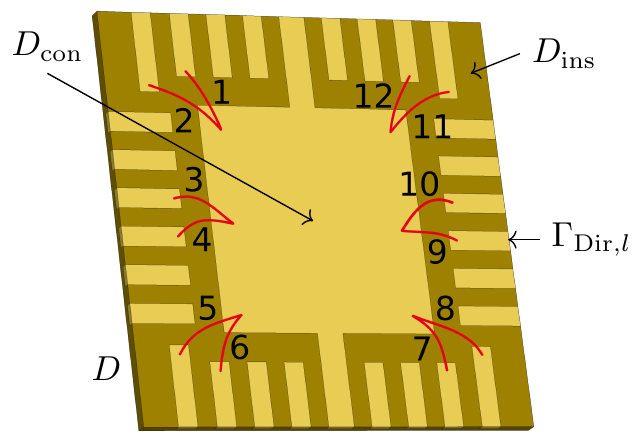}
     \caption{Computational domain $D$ with boundary $\partial D$ consisting of a microelectronic chip package with bond wires. The package is modeled with a homogeneous conducting domain $D_{\mathrm{con}}$ (chip in the center and contact pads located around the chip) and a weak conducting background domain $D_{\mathrm{ins}}$. The boundary part of a contact pad with index $l$ is denoted $\Gamma_{\mathrm{Dir},l}\subset\partial D$.}
 \label{fig:chip_package}
\end{figure}

The present study is motivated from microelectronic packaging in power electronic applications, see Figure~\ref{fig:chip_package}.
An accurate and detailed thermal design is becoming increasingly important due to smaller chip sizes and, hence, increasing power densities.
Therefore, an accurate electrothermal simulation is a more and more indispensable tool during design.
The simulation requires the coupling of the transient heat equation with an electrokinetic problem through Joule\xspace heating effects and through temperature dependent material parameters.
A challenge arising in these types of applications is the presence of thin wires used within the bonding process, i.e., the different scales of the wire and the surrounding package.
The approach adapted here, which is not restricted to microelectronics, consists of modeling wires as 1D structures and computing the potential, resp.\ temperature, distribution using an additional 1D equation.
The electric current, resp.\ heat flow, to the surrounding is then achieved through a coupling of the 1D substructure with the 3D geometry.
Such singular substructures are frequently encountered and dealt with in electromagnetics \cite{Umashankar_1987aa,Noda_2002aa,Wu_2008aa}.
Here, we present a more detailed mathematical modeling of the problem both on the continuous as well as on the discrete level.
An appropriate tool to model singularities in electromagnetics are de Rham\xspace currents \cite{De-Rham_1984aa} and their discrete counterparts \cite{Auchmann_2007aa}.
We further show that the problem is closely related to fluid flow in porous media with fractures as studied, e.g., in \cite{DAngelo_2007aa,Cattaneo_2014aa}.
Hence, mathematical tools for elliptic problems with Dirac measures \cite{DAngelo_2008aa} can be applied.  

The study was motivated by the analysis of micro- and nanoelectronic problems involving bond wire simulations in the scope of the EU FP-7 project nanoCOPS \cite{ter-Maten_2016ab,Casper_2016aa,Casper_2016ah,Duque_2017ab}.
We use the \gls{FIT} on a pair of rectilinear grids as proposed in~\cite{Weiland_1977aa,Weiland_1996aa,Clemens_2001ac}.
Its use is motivated by the underlying rectilinear structure of typical chip package geometries.
In this paper, due to the 1D-3D coupling, the curved geometry of a wire can be incorporated without suffering from a staircase approximation.
Staircase approximations are very common for thin wire approximations in electromagnetics and a discussion of the associated drawbacks can be found in \cite{Noda_2004aa}.
From a \gls{FE} context, the \gls{FIT} can be understood as an \gls{FE} method on a hexahedral grid where an appropriate mass lumping is applied to the material matrices~\cite{Bossavit_2000ab,Bossavit_1999af,Bondeson_2005aa}.
In addition to Whitney\xspace \gls{FE}, strong relations exist to other numerical schemes, such as mimetic finite differences, see, e.g.\,~\cite{Brezzi_2006aa,Vandekerckhove_2014aa}, allowing for an enhanced grid element variety. 

The paper is structured as follows: in Section~\ref{sec:continuous_system}, we present the electrothermal coupled problem with the additional 1D-3D coupling.
The numerical discretization is addressed in Section~\ref{sec:discrete_system} together with a detailed discussion of boundary conditions.
In Section~\ref{sec:error}, we relate the scheme to a \gls{FE} method for flow problems with one-dimensional fissures.
Finally, a numerical convergence rate study for a simplified elliptic model problem and the transient electrothermal simulation of a microelectronic chip package is presented in Section~\ref{sec:numerics}.
 \section{Continuous Electrothermal Problem}
\label{sec:continuous_system}
The computational domain $D$ of the considered application consists of a microelectronic chip package with $\overline{N}$ applied bond wires as depicted in Figure~\ref{fig:chip_package}.
The subdomain modeled by conducting parts, such as the chip in the middle of the domain and the contact pads, is denoted with $D_{\mathrm{con}}$, whereas $D_{\mathrm{ins}}$ denotes the insulating subdomain.
Electrical signals are imposed on contact electrodes $\Gamma_{\mathrm{Dir},i}\subset\partial D$ and transferred to the chip by bond wires.
For a concise notation, we assume that only one wire is present and release the need for an additional index.
Due to the typically very small radius of a wire compared to the surrounding geometry, we model a wire as the 3D curve $\Lambda$ with zero radius.
Furthermore, we establish a 1D domain $\overline{\Lambda}$ by using a wire parametrization that maps from $\overline{\Lambda}$ to the 3D curve $\Lambda$.
To distinguish quantities associated to the 3D domain $D$ from those associated to the 1D domain $\overline{\Lambda}$, we denote the latter by an overline.
In the following, a formulation for this coupling is presented and we consider the general case of a wire and refer to bond wires as a specific application.

\subsection{Strong Formulation}

Let $I:=(0,t_{0}]$ be the time interval of interest.
We consider capacitive effects only for the thermal 3D part and examine the coupling of an electrokinetic problem with the transient heat equation.
Due to the small wire radii, we assume that the resistive losses in the wires are dominating and we thus neglect the losses in the 3D domain.
Here, we use the Dirac\xspace distribution $\delta_{\Lambda}$ to formulate the coupled 1D-3D electrothermal problem as 
\begin{subequations}
    \begin{alignat}{3}
        -\nabla \cdot\left(\sigma(\mathbf{x},T)\nabla\varphi(\mathbf{x},t)\right)                                                         &=q_{\sigma}(\mathbf{x},\overline{\varphi})\delta_{\Lambda},                      &&\mathbf{x}&&\in D,t\in I,               \label{eq:electric3D}\\
        -\frac{\partial}{\partial s}\left(\overline{\sigma}\left(s,\overline{T}\right)\frac{\partial}{\partial s}\overline{\varphi}\left(s,t\right)\right)&=\overline{q}_{\sigma}(s,\varphi),                                       && s&&\in\overline{\Lambda},t\in I,\\
        \rho(\mathbf{x})c(\mathbf{x})\partial_{t}T(\mathbf{x},t)-\nabla \cdot\left(\lambda(\mathbf{x},T)\nabla T(\mathbf{x},t)\right)                             &=q_{\lambda}(\mathbf{x},\overline{T})\delta_{\Lambda},                           &&\mathbf{x}&&\in D,t\in I,               \label{eq:thermal3D}\\
        -\frac{\partial}{\partial s}\left(\overline{\lambda}(s,\overline{T})\frac{\partial}{\partial s}\overline{T}(s,t)\right)                           &=\overline{q}_{\lambda}(s,T)+\overline{Q}^{\mathrm{w}}(\overline{\varphi}),\quad&& s&&\in\overline{\Lambda},t\in I,\\
        \overline{\varphi}(s,t)                                                                                   &=\Pi\varphi(\mathbf{x},t),                                                    &&\mathbf{x}&&\in D,s\in\overline{\Lambda},t\in I,\label{eq:electricCoupling}\\
        \overline{T}(s,t)                                                                                         &=\Pi T(\mathbf{x},t),                                                         &&\mathbf{x}&&\in D,s\in\overline{\Lambda},t\in I,\label{eq:thermalCoupling}
\intertext{for the 3D and 1D electric potentials $\varphi$ and $\overline{\varphi}$ as well as the temperatures $T$ and $\overline{T}$, respectively, where we omit the dependency of $\varphi$ on $T$ and vice versa.
Furthermore, all materials are regarded as time-invariant, the 3D electric and thermal conductivities are given by $\sigma$ and $\lambda$, respectively, $\rho$ is the volumetric mass density and $c$ is the specific heat capacity.
On the other hand, the 1D wire material coefficients are given by $\overline{\sigma}=|\overline{A}|\sigma^{\mathrm{w}}$ and $\overline{\lambda}=|\overline{A}|\lambda^{\mathrm{w}}$, where $|\overline{A}|$ is the wire's physical cross-sectional area and $\sigma^{\mathrm{w}}$ and $\lambda^{\mathrm{w}}$ are the electric and thermal conductivities of the wire's material.
The electrothermal coupling is established by the temperature dependent materials and the 1D resistive losses $\overline{Q}^{\mathrm{w}}$.
While the conductivities may depend on temperature, the temperature dependence of $\rho$ and $c$ is neglected.
The 1D-3D formulation requires a coupling for the currents (thermal flows) and for the electric potential (temperatures).
The former is denoted by the contributions $q_{\sigma}$ ($q_{\lambda}$) from 1D to 3D domain and by $\overline{q}_{\sigma}$ ($\overline{q}_{\lambda}$) from 3D to 1D domain.
The latter is given by the coupling operator $\Pi$ that will be defined in Section~\ref{subsec:couplingOperator}.
Note that no explicit boundary conditions for the 1D case are required due to the strong coupling by $\Pi$.
On the other hand, the 3D boundary conditions are given by
}
        \varphi(\mathbf{x},t)                                                                                            &=\Phi_{i},                                                              &&\mathbf{x}&&\in\Gamma_{\mathrm{Dir},i},t\in I,\\
        \vec{n}_{D}\cdot(\sigma(\mathbf{x},T)\nabla\varphi(\mathbf{x},t))                                                   &=0,                                                                     &&\mathbf{x}&&\in\Gamma_{\mathrm{Neu}},t\in I,\\
        -\vec{n}_{D}\cdot(\lambda(\mathbf{x},T)\nabla T(\mathbf{x},t))                                                     &=r(T)-r(T_{\infty}),                                                    &&\mathbf{x}&&\in\partial D,t\in I        \label{eq:electrothermal5}\intertext{where the Dirichlet\xspace (Neumann\xspace) part of the boundary is denoted by $\Gamma_{\mathrm{Dir},i}$ ($\Gamma_{\mathrm{Neu}}$), $\Phi_{i}$ is the constant Dirichlet\xspace potential on $\Gamma_{\mathrm{Dir},i}$ and $\vec{n}_{D}$ is the outer unit normal of $D$.
Convective heat exchange with the environment is modeled by the Robin\xspace boundary condition \eqref{eq:electrothermal5}, with the ambient temperature $T_{\infty}$ and $r(T)=hT$, where $h$ denotes the heat transfer coefficient.
Although omitted throughout this paper, combined convection with radiation could be imposed by setting $r(T)=hT+\varepsilon\sigma_{\mathrm{SB}}T^{4}$, where $\varepsilon$ denotes the emissivity and $\sigma_{\mathrm{SB}}$ the Stefan\xspace-Boltzmann\xspace constant.
The initial conditions read
}
        \varphi(\mathbf{x},0)                                                                                            &=\varphi_{\text{init}},                                                 &&\mathbf{x}&&\in D,\\
        \overline{\varphi}(s,0)                                                                                   &=\Pi\varphi_{\text{init}},                                            && s&&\in\overline{\Lambda},\\
        T(\mathbf{x},0)                                                                                                  &=T_{\text{init}},                                                       &&\mathbf{x}&&\in D,\\
        \overline{T}(s,0)                                                                                         &=\Pi T_{\text{init}},                                                 && s&&\in\overline{\Lambda},
    \end{alignat}
	\label{eq:electrothermal}\end{subequations}
where $\varphi_{\text{init}}$ is the initial potential and $T_{\text{init}}$ the initial temperature.
Note that the initial values $\varphi_{\text{init}}$ and $T_{\text{init}}$ are used for both the 1D and the 3D case.

\subsection{de Rham\xspace Currents}
\label{subsec:deRham}
In electromagnetics, singular sources, such as line and surface currents, point and surface charges are appropriately represented using de Rham\xspace currents~\cite{De-Rham_1984aa}, i.e., distributions on differential forms.
In particular, for the inclusion of wires, line currents are introduced here.
We largely avoid the formalism of exterior calculus and instead, identify differential forms with their vector proxies.
The reader is referred, e.g., to~\cite{De-Rham_1984aa,Cessenat_1996aa} for further details on forms and currents. 

Let $\mathcal{D}_{0}^{p}(D)$ refer to the space of smooth differential $p$-forms. In particular, we can identify $\mathcal{D}_{0}^{0}(D)$ with $C_{0}^{\infty}(D)$ and $\mathcal{D}_{0}^{1}(D)$ with $(C_{0}^{\infty}(D))^{3}$, respectively.
A de Rham\xspace $p$-current is a map from $\mathcal{D}_{0}^{p}(D)$ into the real numbers.
It can be defined by both vector fields and oriented manifolds.
For instance, electric currents such as the current density as well as surface and line currents give rise to $1$-currents.
On the other hand, quantities as the energy density are represented by $0$-currents, see~\cite{Tucker_2009aa}.
A power density $Q$ in the domain $D$ gives rise to a $0$-current
\begin{equation}
    \pazocal{Q}(v)=\int_{D}Qv\ \mathrm{d}\mathbf{x},\quad v\in C_{0}^{\infty}(D).
    \label{eq:zeroCurrent}
\end{equation}
The current density $\vec{J}_{\sigma}:=-\sigma(\cdot,T)\nabla\varphi$ and heat flux density $\vec{J}_{\lambda}:=-\lambda(\cdot,T)\nabla T$, more generally $\vec{J}_{\alpha}$ with $\alpha\in\{\sigma,\lambda\}$ from now on, give rise to a $1$-current
\begin{equation}
    \pazocal{J}_{\alpha}(\vec{v}):=\int_{D}\vec{J}_{\alpha}\cdot\vec{v}\ \mathrm{d}\mathbf{x},\quad\vec{v}\in(C_{0}^{\infty}(D))^3.
    \label{eq:regularCurrent}
\end{equation}
An oriented curve $\Lambda$ with unit tangent vector $\vec{t}$ gives rise to a $1$-current
\begin{equation}
    \pazocal{I}_{\alpha}(\vec{v}):=\int_{\Lambda}I_{\alpha}\vec{t}\cdot\vec{v}\ \mathrm{d}\mathbf{x},\quad\vec{v}\in(C_{0}^{\infty}(D))^3,
\label{eq:lineCurrent}
\end{equation}
with a line current $I_{\alpha}(\mathbf{x}),\mathbf{x}\in\Lambda$.
This current may vary along the curve $\Lambda$ since we allow a current exchange between the curve and its surrounding.
The divergence of any $1$-current $\pazocal{K}$ is defined as~\cite{Auchmann_2007aa}
\begin{equation}
    \dddiv\pazocal{K}(v)=-\pazocal{K}(\nabla v),\quad\forall v\in C_{0}^{\infty}(D). 
    \label{eq:div1current}
\end{equation}
To illustrate how \eqref{eq:electric3D} and \eqref{eq:thermal3D} can be reformulated in the setting of de Rham\xspace currents, we multiply these equations with $v\in C_{0}^{\infty}(D)$, integrate over $D$ and integrate by parts to obtain
\begin{equation}
    -\int_{D}\nabla\cdot(\alpha(\cdot,T)\nabla\varphi)v\ \mathrm{d}\mathbf{x}=\int_{D}\nabla\cdot\vec{J}_{\alpha}v\ \mathrm{d}\mathbf{x}=-\int_{D}\vec{J}_{\alpha}\cdot\nabla v\ \mathrm{d}\mathbf{x}=\dddiv\pazocal{J}_{\alpha}(v)=0,
\end{equation}
in the sense of distributions.
Hence, the de Rham\xspace current reformulation of \eqref{eq:electric3D} without the wire contribution reads
\begin{equation}
    \dddiv\pazocal{J}_{\sigma}(v)=0,\quad\forall v\in C_{0}^{\infty}(D).
    \label{eq:dist}
\end{equation}

Furthermore, by the definition of~\eqref{eq:zeroCurrent}, the de Rham\xspace $0$-current associated to $\rho c\,\partial_{t}T$ reads 
\begin{equation}
    \dot{\pazocal{Q}}_{\rho c}(v)=\int_{D}\rho c\partial_{t}T v\ \mathrm{d}\mathbf{x},\quad\forall v\in C_{0}^{\infty}(D).
\end{equation}
Then, the de Rham\xspace current reformulation of \eqref{eq:thermal3D} without the wire contribution reads
\begin{equation}
    \dot{\pazocal{Q}}_{\rho c}(v)+\dddiv\pazocal{J}_{\lambda}(v)=\pazocal{Q}(v),\quad\forall v\in C_{0}^{\infty}(D).
    \label{eq:current_equation_3d}\end{equation}

From now on, we account for the presence of $\overline{N}$ wires by an additional index $i$.
The embedding of a 1D wire current into the 3D domain is achieved by adding a line current to~\eqref{eq:dist} and~\eqref{eq:current_equation_3d}.
To this end, let the $i$-th wire curve be parametrized as ${\Lambda^{i}=\{\mathbf{x}^{i}(s),s\in\overline{\Lambda}=(0,1)\}}$, such that all one-dimensional quantities can be defined on the interval $[0,1]$.

The embedding into the 3D domain is achieved by considering the image of the map $\mathbf{x}^{i}$ of the 1D current $\overline{I}_{\sigma}^{i}:=-\overline{\sigma}^{i}\partial_{s}\overline{\varphi}^{i}$ (heat flow $\overline{I}_{\lambda}^{i}:=-\overline{\lambda}^{i}\partial_{s}\overline{T}^{i}$) and the associated de Rham\xspace current $\overline{\pazocal{I}}_{\sigma}^{i}$ ($\overline{\pazocal{I}}_{\lambda}^{i}$)~\cite[p.47]{De-Rham_1984aa}.
More precisely, again using $\alpha=\{\sigma,\lambda\}$ for compact notation, we set 
\begin{equation}
    \pazocal{I}_{\alpha}^{i}(\vec{v})=\overline{\pazocal{I}}_{\alpha}^{i}((\mathbf{x}^{i})^{*}\vec{v})
    =\int_{\overline{\Lambda}}\overline{I}_{\alpha}^{i}\left(\frac{\mathrm{d}}{\mathrm{d}s}\mathbf{x}^{i}\right)\cdot\left(\vec{v}\circ\mathbf{x}^{i}\right)\ \mathrm{d} s,\quad\forall\vec{v}\in(C_{0}^{\infty}(D))^{3},
    \label{eq:wireNonDiv}
\end{equation}
where $(\mathbf{x}^{i})^{*}$ denotes the pullback by the map $\mathbf{x}^{i}$.
After applying \eqref{eq:div1current} to \eqref{eq:wireNonDiv},
\begin{align}
    \dddiv\pazocal{I}^{i}_{\alpha}(v)=-\int_{\overline{\Lambda}}\overline{I}_{\alpha}^{i}\left(\frac{\mathrm{d}}{\mathrm{d}s}\mathbf{x}^{i}\right)\cdot\left(\nabla v\circ\mathbf{x}^{i}\right)\ \mathrm{d} s,\quad\forall v\in C_{0}^{\infty}(D),
    \label{eq:wire}
\end{align}
and the total electric current and thermal heat flow wire contribution is obtained from the single wire $1$-currents as $\pazocal{I}_{\alpha}:=\sum_{i=1}^{\overline{N}}\pazocal{I}^{i}_{\alpha}$.

Using~\eqref{eq:zeroCurrent}, the wire Joule\xspace losses can be expressed as
\begin{equation}
    \pazocal{Q}^{\mathrm{w},i}(v)=\overline{\pazocal{Q}}^{\mathrm{w},i}((\mathbf{x}^{i})^{*}v)
    =\int_{\overline{\Lambda}}\overline{Q}^{\mathrm{w},i}v\circ\mathbf{x}^{i}\ \mathrm{d} s,\quad\forall v\in C_{0}^{\infty}(D),
    \label{eq:wire_losses}
\end{equation}
where $\overline{Q}^{\mathrm{w},i}:=\overline{\sigma}^{i}\left|\partial_{s}\overline{\varphi}^{i}\right|^{2}$.
Again, the overall wire contribution $\pazocal{Q}^{\mathrm{w}}$ is obtained by summing over all wires. 

With these definitions at hand, by adding~\eqref{eq:wire} and~\eqref{eq:wire_losses} to~\eqref{eq:dist} and~\eqref{eq:current_equation_3d}, the 1D-3D coupled problem in terms of de Rham\xspace currents is obtained as
\begin{subequations}
    \begin{align}
                                      \dddiv\left(\pazocal{J}_{\sigma}+\pazocal{I}_{\sigma}\right)(v)&=0,\\
        \dot{\pazocal{Q}}_{\rho c}(v)+\dddiv\left(\pazocal{J}_{\lambda}+\pazocal{I}_{\lambda}\right)(v)&=\pazocal{Q}^{\mathrm{w}}(v),\label{eq:current_equation2}
    \end{align}
    \label{eq:current_equation}\end{subequations}
for all $v\in C_{0}^{\infty}(D)$.

\subsection{Coupling Operator}
\label{subsec:couplingOperator}
\begin{figure}
    \centering
    \includegraphics{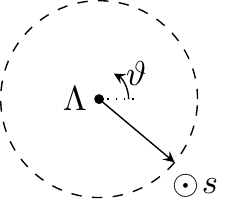}
    \caption{Cross-section of a wire with its radius confined to a 1D line $\overline{\Lambda}$ in $s$-direction. Further denoted are the coupling radius $r_{\mathrm{cpl}}$ and the angle $\vartheta$ around the wire.}
    \label{fig:wireCrossSection}
\end{figure}

What is left is the definition of the coupling operator $\Pi$ as used in~\eqref{eq:electricCoupling} and~\eqref{eq:thermalCoupling}.
Due to the singularity of the 3D solution at $\Lambda$, a direct coupling of 3D and 1D solution at $\Lambda$ is not possible.
Instead, following~\cite{DAngelo_2008aa}, we use an averaging scheme given by
\begin{equation}
    \Pi u:=\gamma\frac{1}{2\pi}\int_{0}^{2\pi}u(\mathbf{x}(s,r_{\mathrm{cpl}},\vartheta))\ \mathrm{d}\vartheta,
    \label{eq:cplCondition}
\end{equation}
where $u\in\{\varphi,T\}$ and $\gamma$ is a scaling coefficient.
Note that $(s,r_{\mathrm{cpl}},\vartheta)$ refer to cylindrical coordinates around $\overline{\Lambda}$, where $r_{\mathrm{cpl}}$ is the coupling radius, see Figure~\ref{fig:wireCrossSection}.
Typically, $r_{\mathrm{cpl}}\gg\overline{r}$, where $\overline{r}$ is the radius of the cylindrical wire, is chosen to circumvent resolving the wire.

For a 3D diffusion problem  with Dirac\xspace right hand side as given by~\eqref{eq:electric3D} and~\eqref{eq:thermal3D}, cylindrical coordinates, an infinite domain $D$ and $\Lambda$ coinciding with the $z$-axis, the solution has the form
\begin{align}
    u(\mathbf{x}(s,r,\cdot))=-\frac{q_{\alpha}(\mathbf{x}(s,0,\cdot),\overline{u})\left|\overline{A}\right|}{2\pi\alpha}\log\left(\frac{r}{r_{0}}\right),
    \label{eq:solutionLineSource}
\end{align}
where $r_{0}$ is a reference radius.
With this logarithmic solution, ${\gamma=\log(\overline{r}/r_{0})/\log(r_{\mathrm{cpl}}/r_{0})}$ scales from the averaged value to the physical value of the line source solution at the wire radius $\overline{r}$.
With this definition of $\gamma$, applying~\eqref{eq:cplCondition} to~\eqref{eq:solutionLineSource} yields the corresponding 1D solution given by
\begin{align}
    \overline{u}(s)=-\frac{q_{\alpha}(\mathbf{x}(s,0,\cdot),\overline{u})\left|\overline{A}\right|}{2\pi\alpha}\log\left(\frac{\overline{r}}{r_{0}}\right).
    \label{eq:solution1D}
\end{align}
Finally, for arbitrary curves $\Lambda$, the above solutions are valid in a sufficiently small neighborhood of $\Lambda$. 

\section{Discrete Electrothermal Problem}
\label{sec:discrete_system}
Discretization of the electrothermal problem is carried out using the \gls{FIT}~\cite{Weiland_1977aa,Weiland_1996aa,Clemens_2001ac} motivated by its natural relation to discrete differential forms~\cite{Clemens_2001aa}.
We emphasize that no conceptual difficulty would arise when a comparable discretization scheme as, e.g., \gls{FE}, would be used instead.
A more detailed discussion of this aspect and the numerical discretization errors are given in Section~\ref{sec:error}.
In this section, we define discrete currents in analogy to the continuous de Rham\xspace currents.
Then, we successively introduce the 3D and 1D discretizations.
Further, the averaging of the materials requires a dual grid that is presented in Section~\ref{subsec:dual}.
Finally, in Section~\ref{subsec:boundary}, boundary conditions are formulated and included in the discrete system of equations.

For the discretization, we choose standard lowest order nodal functions on a rectilinear grid, also referred to as nodal Whitney\xspace functions.
Whitney\xspace functions are the discrete counterpart of differential forms and hence, discrete currents are defined as maps from the space of Whitney\xspace functions into the real numbers.
In this way we obtain the discrete counterpart of \eqref{eq:current_equation} (denoting discrete de Rham\xspace currents with the same symbols as their continuous counterparts) as
\begin{subequations}
\begin{align}
                                            \dddiv\left(\pazocal{J}_{\sigma}   +\pazocal{I}_{\sigma}   \right)(\mathpzc{E}_{k})&=0,\label{eq:current_discrete1}\\
\dot{\pazocal{Q}}_{\rho c}(\mathpzc{E}_{k})+\dddiv\left(\pazocal{J}_{\lambda}+\pazocal{I}_{\lambda}\right)(\mathpzc{E}_{k})&=\pazocal{Q}^{\mathrm{w}}(\mathpzc{E}_{k}), \label{eq:current_discrete2}
\end{align}
\label{eq:current_discrete}\end{subequations}
where $\mathpzc{E}_{k}$ refers to the nodal Whitney\xspace function associated with node $k$.
In the following, we will derive the expressions that are required for the implementation of \eqref{eq:current_discrete}.

\subsection{3D Discretization}
\label{sec:3Ddiscrete}
For the discretization, a rectilinear grid with $N^{\text{N}}$ nodes, $N^{\text{E}}$ edges, $N^{\text{F}}$ facets and $N^{\text{C}}$ cells is used.
With $k=1,\dots,N^{\text{N}}\text{ and }l=1,\dots,N^{\text{E}}$, we denote with $P_{k}$ and $L_{l}$ the nodes (points) and edges (lines) of the grid, respectively.
In this subsection, we omit the wire contribution.
The standard discrete gradient operator is introduced as the primal node to edge incidence matrix $\G\in\{-1,0,1\}^{N^{\text{E}}\times N^{\text{N}}}$ as
\begin{align} 
    \left(\G\right)_{lk}&:=\left\{
    \begin{aligned}
        &1, &&\text{if }P_{k}\in   \overline{L_{l}}\text{ and }P_{k}\text{ is the end point of }     L_{l},\\
        &-1,&&\text{if }P_{k}\in   \overline{L_{l}}\text{ and }P_{k}\text{ is the start point of }L_{l},\\
        &0, &&\text{if }P_{k}\notin\overline{L_{l}}.
    \end{aligned}
    \right.
\end{align}
The discrete counterpart of \eqref{eq:div1current} at the grid level reads
\begin{equation}
    \dddiv\pazocal{K}(\mathpzc{E}_{k})=-\pazocal{K}(\nabla\mathpzc{E}_{k})+BC,
    \label{eq:discrete_stokes}
\end{equation}
for any discrete 1-current $\pazocal{K}$ and possible boundary contributions $BC$, see~\cite{Auchmann_2007aa}.
This boundary contribution is omitted for the time being and is discussed in Section \ref{subsec:boundary}.
Thus, by applying \eqref{eq:discrete_stokes} and evaluating the Whitney\xspace functions on the grid, we obtain
\begin{equation}
    \dddiv\pazocal{J}_{\alpha}(\mathpzc{E}_{k})=-\mathbf{j}_{\alpha}^{\top}\G\mathbf{e}_{k}=-(\G^{\top}\mathbf{j}_{\alpha})_{k},
    \label{eq:div1current_discrete}
\end{equation}
with the coefficient vector $\mathbf{j}_{\alpha}$ of the current $\vec{J}_{\alpha}$, see again~\cite{Auchmann_2007aa}, and the unit basis vector $\mathbf{e}_{k}$.
Applying \eqref{eq:div1current_discrete} to \eqref{eq:current_discrete}, we obtain the discrete system
\begin{subequations}
    \begin{align}
                         -\G^{\top}\mathbf{j}_{\sigma} (\ensuremath{\boldsymbol{\mathrm{\varphi}}})&=\ensuremath{\mathbf{0}},\\
        \dot{\Q}_{\rho c}-\G^{\top}\mathbf{j}_{\lambda}(\ensuremath{\mathbf{T}})   &=\ensuremath{\mathbf{0}},
    \end{align}
    \label{eq:electrothermal_discrete_fluxes}\end{subequations}
where $\dot{\Q}_{\rho c}$ denotes the coefficient vector of the current $\dot{Q}_{\rho c}$.
The potential and temperature evaluated at the grid points are the degrees of freedom given by $\ensuremath{\boldsymbol{\mathrm{\varphi}}}\in\mathbb{R}^{N^{\text{N}}}$ and $\ensuremath{\mathbf{T}}\in\mathbb{R}^{N^{\text{N}}}$, respectively. 

\subsection{1D Discretization and Coupling to 3D Domain}
\begin{figure}[t]
 \centering
 \includegraphics{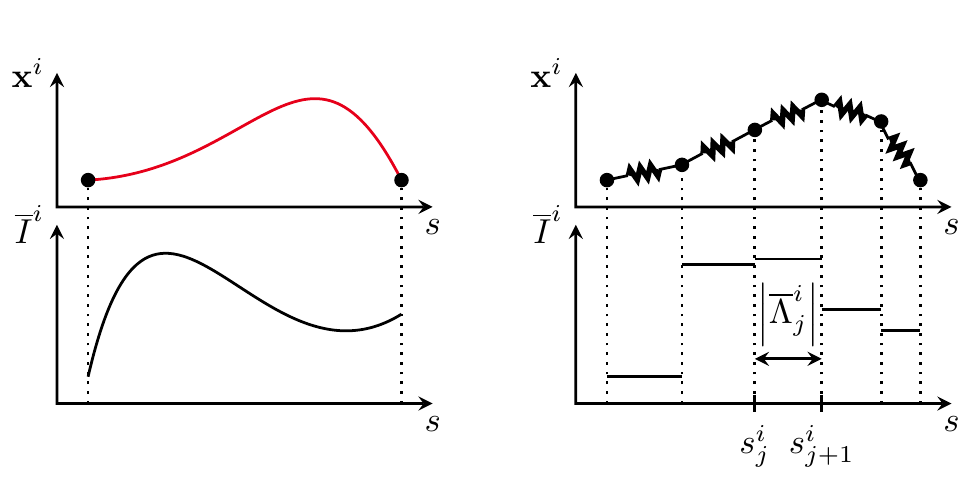}
 \caption{1D discretization of wire $i$ visualized using lumped elements for the example of $N^{\mathrm{1D},i}=6$ points with thermal and electrical resistances. The wire's current is discretized using piecewise constant basis functions.}
 \label{fig:wire_lumped}
\end{figure}

\begin{figure}[t]
 \centering
 \includegraphics{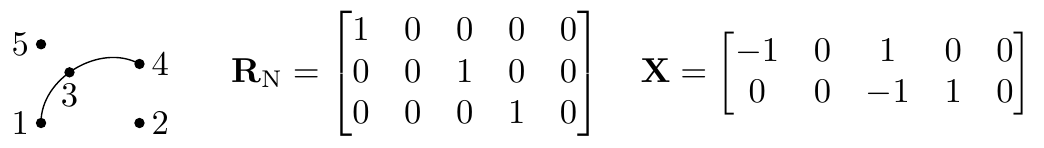}
 \caption{An exemplary wire located among five arbitrary grid points of which points 1, 3 and 4 coincide with 1D grid points. The corresponding coupling matrices $\ensuremath{ \mathbf{R} }_{\text{N}}$ and $\ensuremath{ \mathbf{X} }$ are shown.}
 \label{fig:visualizeX}
\end{figure}

The \gls{FIT}-discretization of $\overline{I}_{\alpha}$ results in a piecewise constant discretization on every wire part (curved element) $\Lambda_{j}^{i}$ for wire $i=1,\ldots,\overline{N}$.
We model each wire part as a 1D element giving a series connection of $N^{\mathrm{1D},i}-1$ elements for wire $i$, cf. Figure~\ref{fig:wire_lumped}.
We obtain the aforementioned partitioning of $\overline{\Lambda}^{i}$ by dividing $[0,1]$ with nodes $\{s_{j}^{i}\}$, $j=1,\ldots,N^{\mathrm{1D},i}$.
Let ${\overline{\ensuremath{ \mathbf{P} }}_{s}^{i}\in\{-1,0,1\}^{(N^{\mathrm{1D},i}-1)\times N^{\mathrm{1D},i}}}$ denote the partial derivative operator of wire $i$, defined as
\begin{align} 
    \left(\overline{\ensuremath{ \mathbf{P} }}_{s}^{i}\right)_{lk}&:=\left\{
    \begin{aligned}
        &1, &&\text{if }k=l+1,\\
        &-1,&&\text{if }k=l,\\
        &0, &&\text{otherwise}.
    \end{aligned}
    \right.
\end{align}
Furthermore, let 
\begin{align}
    \overline{\ensuremath{ \mathbf{M} }}_{\alpha}^{i}:=\text{diag}\left(\frac{\overline{\alpha}_{1}^{i}}{|\Lambda_{1}^{i}|},\ldots,\frac{\overline{\alpha}_{N^{\mathrm{1D},i}-1}^{i}}{|\Lambda_{N^{\mathrm{1D},i}-1}^{i}|}\right),
\end{align}
represent the electrical/thermal 1D wire mass matrix.
Then, the \gls{FIT} discretization of the 1D current and heat flow reads
\begin{subequations}
    \begin{align}
    \overline{\mathbf{I}}_{\sigma}^{i} &=-\overline{\ensuremath{ \mathbf{M} }}_{\sigma}^{i} \overline{\ensuremath{ \mathbf{P} }}_{s}^{i}\overline{\ensuremath{\boldsymbol{\mathrm{\varphi}}}}^{i},\\
    \overline{\mathbf{I}}_{\lambda}^{i}&=-\overline{\ensuremath{ \mathbf{M} }}_{\lambda}^{i}\overline{\ensuremath{ \mathbf{P} }}_{s}^{i}\overline{\ensuremath{\mathbf{T}}}^{i}.
    \end{align}
    \label{eq:1D_equation_discrete}\end{subequations}
We proceed by discretizing \eqref{eq:wire} as
\begin{equation}
    \begin{aligned}
        \dddiv\pazocal{I}_{\alpha}^{i}(\mathpzc{E}_{k})
        &=-\sum_{j=1}^{N^{\mathrm{1D},i}-1}\overline{I}_{\alpha,j}^{i}\int_{s_{j}^{i}}^{s_{j+1}^{i}}\left(\frac{\mathrm{d}}{\mathrm{d}s}\mathbf{x}^{i}\right)\cdot(\nabla\mathpzc{E}_{k}\circ\mathbf{x}^{i})\ \mathrm{d} s\\
        &=-\sum_{j=1}^{N^{\mathrm{1D},i}-1}\overline{I}_{\alpha,j}^{i}\left(\mathpzc{E}_{k}(\mathbf{x}^{i}(s_{j+1}^{i}))-\mathpzc{E}_{k}(\mathbf{x}^{i}(s_{j}^{i}))\right)\\
        &=-\sum_{j=1}^{N^{\mathrm{1D},i}-1}\overline{I}_{\alpha,j}^{i}\left((\ensuremath{ \mathbf{R} }_{\text{N}}^{i})_{j+1,k}-(\ensuremath{ \mathbf{R} }_{\text{N}}^{i})_{j,k}\right)
         =-((\overline{\ensuremath{ \mathbf{P} }}_{s}^{i}\ensuremath{ \mathbf{R} }_{\text{N}}^{i})^{\top}\overline{\mathbf{I}}_{\alpha}^{i})_{k},
    \end{aligned}
    \label{eq:wire_div}
\end{equation}
where we have introduced the coupling matrix $\ensuremath{ \mathbf{R} }_{\text{N}}^{i}\in\mathbb{R}^{N^{\mathrm{1D},i}\times N^{\text{N}}}$ by $(\ensuremath{ \mathbf{R} }_{\text{N}}^{i})_{j,k}:=\mathpzc{E}_{k}(\mathbf{x}^{i}(s_{j}^{i}))$.
It should be noted that, in view of \eqref{eq:wire_div}, no approximation of the (possibly curved) wire geometry by the underlying 3D grid is required, which is the main advantage of the 1D-3D coupling.
To simplify notation, we further introduce $\ensuremath{ \mathbf{X} }^{i}\in\mathbb{R}^{(N^{\mathrm{1D},i}-1)\times N^{\text{N}}}$ as $\ensuremath{ \mathbf{X} }^{i}:=\overline{\ensuremath{ \mathbf{P} }}_{s}^{i}\ensuremath{ \mathbf{R} }_{\text{N}}^{i}$ (cf. Figure~\ref{fig:visualizeX}) and infer the vector representation for $\dddiv\pazocal{I}_{\sigma}^{i}$, combining \eqref{eq:1D_equation_discrete} and \eqref{eq:wire_div} as
\begin{equation}
    \left(\dddiv\pazocal{I}_{\alpha}^{i}(\mathpzc{E}_{1}),\dots,\dddiv\pazocal{I}_{\alpha}^{i}(\mathpzc{E}_{N^{\text{N}}})\right)^{\top}=(\ensuremath{ \mathbf{X} }^{i})^{\top}\overline{\ensuremath{ \mathbf{M} }}_{\alpha}^{i}\overline{\ensuremath{ \mathbf{P} }}_{s}^{i}\overline{\ensuremath{\boldsymbol{\mathrm{\varphi}}}}^{i}.
\end{equation}
\begin{rmk}
The coupling matrix $\ensuremath{ \mathbf{R} }_{\text{N}}^{i}$ can be understood as the discrete counterpart of the pullback $(\mathbf{x}^{i})^{*}$.
For the simple case when the 1D nodes are obtained by pulling back the 3D grid nodes, $\ensuremath{ \mathbf{R} }_{\text{N}}^{i}\in \{0,1\}^{N^{\mathrm{1D},i}\times N^{\text{N}}}$ becomes an operator that restricts to those nodes of the grid which are connected to wire $i$ and contains only one non-zero entry in each row.
In the general case, each 1D node is allocated in a grid volume defined by \num{8} nodes, this operator becomes $\ensuremath{ \mathbf{R} }_{\text{N}}^{i}\in\mathbb{R}^{N^{\mathrm{1D},i}\times N^{\text{N}}}$ and contains $\num{8}$ non-zero entries in each row, obtained by Whitney\xspace interpolation.
In any case, the sum of all entries in one row is always equal to one.
\end{rmk}

To account for the 1D Joule\xspace losses, we introduce the vector $\overline{\Q}^{\mathrm{w},i}$, allocated at 1D edges, with $N^{\mathrm{1D},i}-1$ entries each given by 
\begin{equation}
    \overline{Q}_{j}^{\mathrm{w},i}=-\overline{I}_{\sigma,j}^{i}(\overline{\ensuremath{ \mathbf{P} }}_{s}^{i}\overline{\ensuremath{\boldsymbol{\mathrm{\varphi}}}}^{i})_{j},
\end{equation}
representing the discretized heat power of wire $i$ with $\overline{I}_{\sigma,j}^{i}$ being the current in element $j$.
Starting from \eqref{eq:wire_losses}, the Joule\xspace losses of the wire part are discretized as 
\begin{equation}
    \begin{aligned}
        \pazocal{Q}^{\mathrm{w}}(\mathpzc{E}_{k})
        &=\sum_{i=1}^{\overline{N}}\sum_{j=1}^{N^{\mathrm{1D},i}-1}\frac{1}{2}\overline{Q}_{j}^{\mathrm{w},i}\left(\mathpzc{E}_{k}(\mathbf{x}^{i}(s_{j+1}^{i}))+\mathpzc{E}_{k}(\mathbf{x}^{i}(s_{j}^{i}))\right)\\
        &=\sum_{i=1}^{\overline{N}}\sum_{j=1}^{N^{\mathrm{1D},i}-1}\frac{1}{2}\left(\ensuremath{ \mathbf{X} }_{\text{abs}}^{i}\right)_{j,k}\overline{Q}_{j}^{\mathrm{w},i}=:(\Q^{\mathrm{w}})_{k},
    \end{aligned}
    \label{eq:Q_wire}
\end{equation}
where the entries of $\ensuremath{ \mathbf{X} }_{\text{abs}}^{i}$ are given by the absolute values of the entries of $\ensuremath{ \mathbf{X} }^{i}$.
Let $\overline{\ensuremath{\mathbf{T}}}_{\text{avg}}^{i}=\frac{1}{2}\ensuremath{ \mathbf{X} }_{\text{abs}}^{i}\ensuremath{\mathbf{T}}$ denote the averaged temperatures for each element of wire $i$. We account for nonlinearities in the wire material parameters as $\alpha_{j}^{i}((\overline{\ensuremath{\mathbf{T}}}_{\text{avg}}^{i})_{j})$.

The last ingredient for the discretization is the discretized version $\boldsymbol{\Pi}\in\mathbb{R}^{N^{\mathrm{1D}}\times N^{\text{N}}}$ of the coupling operator $\Pi$ as introduced in Section~\ref{subsec:couplingOperator}.
It is obtained by interpolating $\mathbf{x}^{*}\Pi u$ at the nodes of the 1D grid to yield the relations
\begin{equation}
    \overline{\ensuremath{\boldsymbol{\mathrm{\varphi}}}}^{i}=\boldsymbol{\Pi}^{i}\ensuremath{\boldsymbol{\mathrm{\varphi}}},\qquad\overline{\ensuremath{\mathbf{T}}}^{i}=\boldsymbol{\Pi}^{i}\ensuremath{\mathbf{T}}
    \label{eq:discrete_projection}
\end{equation} 
for wire $i$. Finally, by introducing the lumped wire stiffness matrix $\ensuremath{\mathbf{K}}_{\alpha}^{\mathrm{w}}$ as
\begin{equation}   
    \ensuremath{\mathbf{K}}_{\alpha}^{\mathrm{w}}:=\sum_{i=1}^{\overline{N}}\ensuremath{\mathbf{K}}_{\alpha}^{\mathrm{w},i}=\sum_{i=1}^{\overline{N}}(\ensuremath{ \mathbf{X} }^{i})^{\top}\overline{\ensuremath{ \mathbf{M} }}_{\alpha}^{i}  \overline{\ensuremath{ \mathbf{P} }}_{s}^{i}\boldsymbol{\Pi}^{i},
\end{equation}
we conclude
\begin{subequations}
    \begin{align}
        \dddiv\pazocal{I}_{\sigma} (\mathpzc{E}_{k})&=(\ensuremath{\mathbf{K}}_{\sigma}^{\mathrm{w}} \ensuremath{\boldsymbol{\mathrm{\varphi}}})_{k},\\
        \dddiv\pazocal{I}_{\lambda}(\mathpzc{E}_{k})&=(\ensuremath{\mathbf{K}}_{\lambda}^{\mathrm{w}}\ensuremath{\mathbf{T}})_{k}.
    \end{align}
    \label{eq:div_wire}\end{subequations}

\subsection{Dual Grid}
\label{subsec:dual}
In this section, we introduce the dual rectilinear grid containing $\widetilde{N}^{\text{N}}=N^{\text{C}}$ points, $\widetilde{N}^{\text{E}}=N^{\text{F}}$ edges, $\widetilde{N}^{\text{F}}=N^{\text{E}}$ facets and $\widetilde{N}^{\text{C}}=N^{\text{N}}$ cells.
We denote by $\ensuremath{\protect\widetilde{A}}_{l}, \widetilde{V}_{k}$ the facets (areas) and cells (volumes) of the dual grid, respectively. 
\begin{figure}
    \centering
    \includegraphics{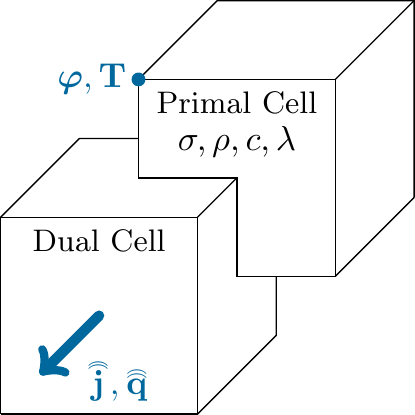}
    \caption{Allocation of electrical and thermal quantities at the primal and dual grid.}
    \label{fig:quantity_allocation}
\end{figure}

The non-singular quantities (i.e.\ not related to wires) can be identified with elements on the dual grid, as commonly done in \gls{FIT} (cf. Figure~\ref{fig:quantity_allocation}).
For instance, the discrete de Rham\xspace current $\pazocal{J}_{\alpha}$ is defined by the coefficients
\begin{align}
    j_{\alpha,l}=\int_{\ensuremath{\protect\widetilde{A}}_{l}}\vec{J}_{\alpha}\cdot\vec{n}_{\ensuremath{\protect\widetilde{A}}_{l}}\ \mathrm{d}\mathbf{x},
\end{align}
with the unit normal $\vec{n}_{\ensuremath{\protect\widetilde{A}}_{l}}$ of facet $\ensuremath{\protect\widetilde{A}}_{l}$.
Also, the discrete de Rham\xspace current $\pazocal{Q}$ is represented by the vector $\Q\in\mathbb{R}^{N^{\text{N}}}$, where 
\begin{align}
    Q_{k}=\int_{\widetilde{V}_{k}}Q\ \mathrm{d}\mathbf{x}.
\end{align}

Applying the standard \gls{FIT} material approximation to the unknown current $\vec{J}_{\alpha}$, we obtain 
\begin{multline}
      \int_{\ensuremath{\protect\widetilde{A}}_{l}}\vec{J}_{\alpha}\cdot\vec{n}_{\ensuremath{\protect\widetilde{A}}_{l}}\ \mathrm{d}\mathbf{x}
    =-\int_{\ensuremath{\protect\widetilde{A}}_{l}}(\alpha\nabla\varphi)\cdot\vec{n}_{\ensuremath{\protect\widetilde{A}}_{l}}\ \mathrm{d}\mathbf{x}
    \approx-\frac{\alpha_{l}^{\mathrm{avg}}|\ensuremath{\protect\widetilde{A}}_{l}|}{|\ensuremath{L}_{l}|}\int_{\ensuremath{L}_{l}}\nabla\varphi\cdot\vec{t}_{l}\ \mathrm{d}\mathbf{x}\\
    =      -\frac{\alpha_{l}^{\mathrm{avg}}|\ensuremath{\protect\widetilde{A}}_{l}|}{|\ensuremath{L}_{l}|}(\varphi_{l,1}-\varphi_{l,0})=-\left(\ensuremath{ \mathbf{M} }_{\alpha}\right)_{ll}(\varphi_{l,1}-\varphi_{l,0}),
\end{multline}
where $\left|\ \cdot\ \right|$ denotes the area or length of the specified geometrical object.
The $\alpha_{l}^{\mathrm{avg}}$ are averaged material values defined on primal edges (or dual facets) that are obtained by a suitable averaging scheme~\cite{Clemens_2001aa}.
Furthermore, $\vec{t}_{l}$ refers to the unit tangent of $\ensuremath{L}_{l}$ and $\varphi_{l,0/1}$ to the potential at the start/end point of $\ensuremath{L}_{l}$, respectively.
This defines the electrical and thermal conductance matrices $\ensuremath{ \mathbf{M} }_{\alpha}\in\mathbb{R}^{N^{\text{E}}\times N^{\text{E}}}$ as
\begin{align}
    \ensuremath{ \mathbf{M} }_{\alpha}:=\text{diag}\left(\frac{\alpha_{1}^{\mathrm{avg}}|\ensuremath{\protect\widetilde{A}}_{1}|}{\left|\ensuremath{L}_{1}\right|},
                                  \frac{\alpha_{2}^{\mathrm{avg}}|\ensuremath{\protect\widetilde{A}}_{2}|}{\left|\ensuremath{L}_{2}\right|},\dots,
                                  \frac{\alpha_{N^{\text{E}}}^{\mathrm{avg}}|\ensuremath{\protect\widetilde{A}}_{N^{\text{E}}}|}{\left|\ensuremath{L}_{N^{\text{E}}}\right|}\right).
    \label{eq:Malpha}
\end{align}
Equivalently, the term for the heat powers gives
\begin{equation}
    \int_{\ensuremath{\protect\widetilde{V}}_{k}}Q\ \mathrm{d}\mathbf{x}=\int_{\ensuremath{\protect\widetilde{V}}_{k}}\rho c\dot{T}\ \mathrm{d}\mathbf{x}\approx{\rho c}_{k}^{\mathrm{avg}}|\ensuremath{\protect\widetilde{V}}_{k}|\dot{T}_{k}=\left(\ensuremath{ \mathbf{M} }_{\rho c}\right)_{kk}\dot{T}_{k},
\end{equation}
where the ${\rho c}_{k}^{\mathrm{avg}}$ are average material values defined on primal nodes (or dual cells) and are obtained by a suitable averaging scheme~\cite{Casper_2016aa}.
The thermal capacitance matrix $\mathbf{M}_{\rho c}\in\mathbb{R}^{N^{\text{N}}\times N^{\text{N}}}$ is defined as
\begin{align}
  \ensuremath{ \mathbf{M} }_{\rho c}:=\text{diag}\left({\rho c}_{1}^{\mathrm{avg}}|\ensuremath{\protect\widetilde{V}}_{1}|,{\rho c}_{2}^{\mathrm{avg}}|\ensuremath{\protect\widetilde{V}}_{2}|,\dots,\rho c_{N^{\text{N}}}^{\mathrm{avg}}|\ensuremath{\protect\widetilde{V}}_{N^{\text{N}}}|\right).
\label{eq:Mrhoc}
\end{align}
Note that $\ensuremath{ \mathbf{M} }_{\alpha}$ and $\ensuremath{ \mathbf{M} }_{\rho c}$ are diagonal matrices with strictly positive entries in the \gls{FIT} approach and sparse positive definite matrices in general, e.g., in a Galerkin\xspace \gls{FE} setting.
Hence, we infer
\begin{subequations}
    \begin{align}
        \mathbf{j}_{\sigma}(\ensuremath{\boldsymbol{\mathrm{\varphi}}})&=-\ensuremath{ \mathbf{M} }_{\sigma} (\ensuremath{\mathbf{T}})\G\ensuremath{\boldsymbol{\mathrm{\varphi}}},\\
        \mathbf{j}_{\lambda}(\ensuremath{\mathbf{T}}) &=-\ensuremath{ \mathbf{M} }_{\lambda}(\ensuremath{\mathbf{T}})\G\ensuremath{\mathbf{T}}.
    \end{align}
    \label{eq:fluxes}\end{subequations}

To lighten the notation, we introduce the stiffness matrix $\ensuremath{\mathbf{K}}_{\alpha}(\ensuremath{\mathbf{T}}):=\G^{\top}\ensuremath{ \mathbf{M} }_{\alpha}(\ensuremath{\mathbf{T}})\G$ and collect the 1D potentials (temperatures) of all wires in a matrix denoted by $\overline{\ensuremath{\boldsymbol{\mathrm{\varphi}}}}\in\mathbb{R}^{\overline{N}\times N^{\mathrm{1D}}}$ (${\overline{\ensuremath{\mathbf{T}}}\in\mathbb{R}^{\overline{N}\times N^{\mathrm{1D}}}}$).
Then, combining \eqref{eq:electrothermal_discrete_fluxes} with \eqref{eq:fluxes}, the discrete electrothermal problem with wire contribution given by \eqref{eq:Q_wire} and \eqref{eq:div_wire} reads
\begin{subequations}
    \begin{align}
                       \ensuremath{\mathbf{K}}_{\sigma}(\ensuremath{\mathbf{T}})\ensuremath{\boldsymbol{\mathrm{\varphi}}} +\ensuremath{\mathbf{K}}_{\sigma} ^{\mathrm{w}}(\overline{\ensuremath{\mathbf{T}}})\ensuremath{\boldsymbol{\mathrm{\varphi}}}&=\ensuremath{\mathbf{0}},\label{eq:ETdiscreteEl}\\
        \ensuremath{\mathbf{M}_{\rho c}}\dot{\ensuremath{\mathbf{T}}}+\ensuremath{\mathbf{K}}_{\lambda}(\ensuremath{\mathbf{T}})\ensuremath{\mathbf{T}}  +\ensuremath{\mathbf{K}}_{\lambda}^{\mathrm{w}}(\overline{\ensuremath{\mathbf{T}}})\ensuremath{\mathbf{T}}  &=\mathbf{Q}^{\mathrm{w}}(\overline{\ensuremath{\boldsymbol{\mathrm{\varphi}}}},\overline{\ensuremath{\mathbf{T}}}).
    \end{align}
    \label{eq:dae_detailed}\end{subequations}
By setting $\widehat{\ensuremath{\mathbf{K}}}_{\alpha}(\ensuremath{\mathbf{T}}):=\ensuremath{\mathbf{K}}_{\alpha}(\ensuremath{\mathbf{T}})+\ensuremath{\mathbf{K}}_{\alpha}^{\mathrm{w}}(\overline{\ensuremath{\mathbf{T}}})$ and $\widehat{\mathbf{Q}}(\ensuremath{\boldsymbol{\mathrm{\varphi}}},\ensuremath{\mathbf{T}}):=\mathbf{Q}^{\mathrm{w}}\left(\overline{\ensuremath{\boldsymbol{\mathrm{\varphi}}}},\overline{\ensuremath{\mathbf{T}}}\right)$, we can write~\eqref{eq:dae_detailed} in a more compact form given by
\begin{subequations}
    \begin{align}
                            \widehat{\ensuremath{\mathbf{K}}}_{\sigma}(\ensuremath{\mathbf{T}})\ensuremath{\boldsymbol{\mathrm{\varphi}}}&=\ensuremath{\mathbf{0}},\label{eq:dae_electric}\\
        \ensuremath{ \mathbf{M} }_{\rho c}\dot{\ensuremath{\mathbf{T}}}+\widehat{\ensuremath{\mathbf{K}}}_{\lambda}(\ensuremath{\mathbf{T}})\ensuremath{\mathbf{T}} &=\widehat{\mathbf{Q}}(\ensuremath{\boldsymbol{\mathrm{\varphi}}},\ensuremath{\mathbf{T}}).
    \end{align}
\label{eq:dae_deterministic}\end{subequations}
System~\eqref{eq:dae_deterministic} is a \gls{DAE} and can be integrated in time by using a suitable integration scheme.
Here, we use the implicit Euler\xspace method together with a fractional step method~\cite{Vijalapura_2005aa}, which reads at time step $n$ with uniform step size $\Delta_{t}$
\begin{subequations}
\begin{align}
                                       \widehat{\ensuremath{\mathbf{K}}}_{\sigma} (\ensuremath{\mathbf{T}}^{n})  \ensuremath{\boldsymbol{\mathrm{\varphi}}}^{n+1}&=\ensuremath{\mathbf{0}},\label{eq:dae_electric_euler}\\
    (\ensuremath{\mathbf{T}}^{n+1}+\ensuremath{\mathbf{T}}^{n})\ensuremath{\mathbf{M}_{\rho c}}/\Delta_{t}+\widehat{\ensuremath{\mathbf{K}}}_{\lambda}(\ensuremath{\mathbf{T}}^{n+1})\ensuremath{\mathbf{T}}^{n+1}  &=\widehat{\mathbf{Q}}(\ensuremath{\boldsymbol{\mathrm{\varphi}}}^{n+1},\ensuremath{\mathbf{T}}^{n+1}).
\end{align}
\label{eq:dae_deterministic_fractional}\end{subequations}
The fractional step method has the same order of accuracy as the implicit Euler\xspace method, however, system \eqref{eq:dae_deterministic_fractional} is decoupled and hence, easier to solve.

\subsection{Boundary Conditions}
\label{subsec:boundary}

\begin{figure}[t]
    \centering
    \begin{subfigure}[b]{0.49\textwidth}
        \centering
        \includegraphics{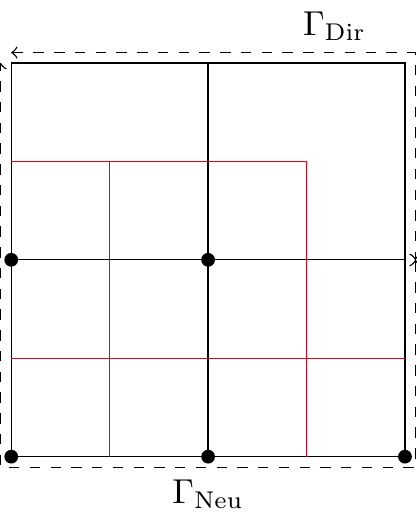}
        \caption{Electrical grid with mixed boundaries}
        \label{fig:electricGrid}
    \end{subfigure}		
    \begin{subfigure}[b]{0.49\textwidth}				
        \centering								
        \includegraphics{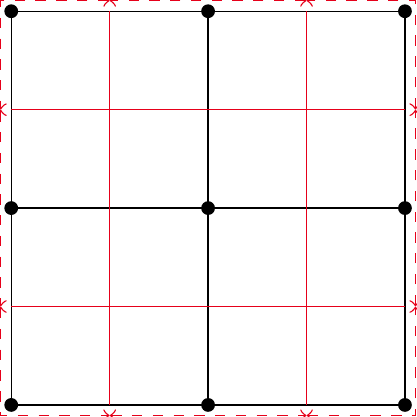}
        \vspace*{1.4em}
        \caption{Thermal augmented grid for Robin boundaries}
        \label{fig:augmentedGrid}
    \end{subfigure}
    \caption{The here used grids are based on~\cite[Fig. 1]{Hiptmair_2001aa}. The primal grid is shown in black, whereas the dual grid is shown in red. Dirichlet\xspace and Neumann\xspace boundaries are indicated by $\Gamma_{\mathrm{Dir}}$ and $\Gamma_{\mathrm{Neu}}$, respectively. Active nodes are shown by the bullets while active boundary facets are indicated by the (red) arrows in (b).}
    \label{fig:grids}
\end{figure}

Boundary conditions are applied for the electrical and thermal subproblem separately.
Starting with the electrical part, to impose Dirichlet\xspace boundary conditions, we follow~\cite{Hiptmair_2001aa} and introduce the notion of active nodes, see~Figure~\ref{fig:electricGrid}.
More precisely, the $N^{N,\mathrm{a}}$-active nodes associated with $\Gamma_{\mathrm{Dir}}$ are obtained by removing all nodes contained in the closure of $\Gamma_{\mathrm{Dir}}$.
Then, with the restriction operators $\ensuremath{ \mathbf{R} }_{\text{N}}^{\mathrm{a}}\in\{0,1\}^{N^{\text{N},\mathrm{a}}\times N^{\text{N}}}$ and $\ensuremath{ \mathbf{R} }_{\text{N}}^{\mathrm{Dir}}\in\{0,1\}^{N^{\text{N},\mathrm{Dir}}\times N^{\text{N}}}$ (denoted as trace operator $\ensuremath{\mathbf{T}}$ in~\cite{Hiptmair_2001aa}), we decompose the electric potential into active nodal potentials $\ensuremath{\boldsymbol{\mathrm{\varphi}}}^{\mathrm{a}}=\ensuremath{ \mathbf{R} }_{\text{N}}^{\mathrm{a}}\ensuremath{\boldsymbol{\mathrm{\varphi}}}$ and prescribed nodal potentials contained in the closure of $\Gamma_{\mathrm{Dir}}$ given by $\ensuremath{\boldsymbol{\mathrm{\varphi}}}^{\mathrm{Dir}}=\ensuremath{ \mathbf{R} }_{\text{N}}^{\mathrm{Dir}}\ensuremath{\boldsymbol{\mathrm{\varphi}}}$.
Based on these definitions, we can reduce~\eqref{eq:dae_electric} as
\begin{equation}
      \ensuremath{ \mathbf{R} }_{\text{N}}^{\mathrm{a}}\widehat{\ensuremath{\mathbf{K}}}_{\sigma}(\ensuremath{\mathbf{T}})(\ensuremath{ \mathbf{R} }_{\text{N}}^{\mathrm{a}}) ^{\top}\ensuremath{\boldsymbol{\mathrm{\varphi}}}^{\mathrm{a}}
    =-\ensuremath{ \mathbf{R} }_{\text{N}}^{\mathrm{a}}\widehat{\ensuremath{\mathbf{K}}}_{\sigma}(\ensuremath{\mathbf{T}})(\ensuremath{ \mathbf{R} }_{\text{N}}^{\mathrm{Dir}})^{\top}\ensuremath{\boldsymbol{\mathrm{\varphi}}}^{\mathrm{Dir}}
\end{equation}
by incorporating Dirichlet\xspace conditions.
To shorten notation, we introduce
\begin{align}
    \widehat{\ensuremath{\mathbf{K}}}_{\sigma}^{\mathrm{a}}(\ensuremath{\mathbf{T}})&:=\ensuremath{ \mathbf{R} }_{\text{N}}^{\mathrm{a}}\widehat{\ensuremath{\mathbf{K}}}_{\sigma}(\ensuremath{\mathbf{T}})(\ensuremath{ \mathbf{R} }_{\text{N}}^{\mathrm{a}}) ^{\top},\\
              \ensuremath{\mathbf{K}}_{\Gamma_\mathrm{Dir}}     &:=\ensuremath{ \mathbf{R} }_{\text{N}}^{\mathrm{a}}\widehat{\ensuremath{\mathbf{K}}}_{\sigma}(\ensuremath{\mathbf{T}})(\ensuremath{ \mathbf{R} }_{\text{N}}^{\mathrm{Dir}})^{\top},
\end{align}
and we obtain the electrical part with boundary conditions
\begin{equation}
    \widehat{\ensuremath{\mathbf{K}}}_{\sigma}^{\mathrm{a}}(\ensuremath{\mathbf{T}})\ensuremath{\boldsymbol{\mathrm{\varphi}}}^{\mathrm{a}}=-\ensuremath{\mathbf{K}}_{\Gamma_{\mathrm{Dir}}}\ensuremath{\boldsymbol{\mathrm{\varphi}}}^{\mathrm{Dir}}.
\end{equation} 

For incorporating the Robin\xspace boundary condition in the thermal equation, it is convenient to consider the augmented dual grid~\cite{Hiptmair_2001aa}, where $N^{\text{N},\partial D}=\widetilde{N}^{\text{F},\partial D}$ dual facets on the boundary are introduced that complete the boundaries of the dual cells, see~Figure~\ref{fig:augmentedGrid}. 
For the augmented dual grid, the boundary contribution $BC$ in \eqref{eq:discrete_stokes} does not vanish. 

Following~\cite{Hiptmair_2001aa}, the additional term is quantified as
\begin{equation}
    BC=(\ensuremath{ \mathbf{R} }_{\text{N}}^{\partial D})^{\top}\widetilde{\ensuremath{ \mathbf{R} }}_{\text{F}}^{\partial D}\mathbf{j}_{\lambda},
\end{equation}
where $\widetilde{\ensuremath{ \mathbf{R} }}_{\text{N}}^{\partial D}\in\{0,1\}^{N^{\text{N},\partial D}\times N^{\text{N}}}$ and $\ensuremath{ \mathbf{R} }_{\text{F}}^{\partial D}\in\{-1,0,1\}^{\widetilde{N}^{\text{F},\partial D}\times\widetilde{N}^{\text{F}}}$ denote the restriction operators to the boundary nodes and the boundary dual facets (of the augmented dual grid), respectively.
Note that the signs in $\widetilde{\ensuremath{ \mathbf{R} }}_{\text{F}}^{\partial D}$ are chosen in such a way that the orientation of the restricted fluxes with respect to the outer normal is taken into account.
Let us define ${\ensuremath{ \mathbf{M} }_{h}^{\partial D}:=\text{diag}(h_{k_{i}}|\partial\ensuremath{\protect\widetilde{V}}_{k_{i}}\cap\partial D|)\in\mathbb{R}^{N^{\text{N},\partial D}\times N^{\text{N},\partial D}}}$, where $k_{i}$ refers to the index of the $i$-th boundary node, ${i=1,\dots,N^{\text{N},\partial D}}$ and let $\ensuremath{\mathbf{T}}_{\infty}$ be a constant vector containing entries $T_{\infty}$. 
Then, by using the discrete Robin\xspace boundary condition
\begin{equation}
    \widetilde{\ensuremath{ \mathbf{R} }}_{\text{F}}^{\partial D}\mathbf{j}_{\lambda}=\ensuremath{ \mathbf{M} }_{h}^{\partial D}\ensuremath{ \mathbf{R} }_{\text{N}}^{\partial D}(\ensuremath{\mathbf{T}}-\ensuremath{\mathbf{T}}_{\infty}),
\end{equation}
and the notation ${\ensuremath{ \mathbf{M} }^{\partial D}:=(\widetilde{\ensuremath{ \mathbf{R} }}_{\text{N}}^{\partial D})^{\top}\ensuremath{ \mathbf{M} }_{h}^{\partial D}\ensuremath{ \mathbf{R} }_{\text{N}}^{\partial D}}$, we obtain the electrothermal system with boundary conditions
\begin{subequations}
    \begin{align}
        \widehat{\ensuremath{\mathbf{K}}}_{\sigma}^{\mathrm{a}}(\ensuremath{\mathbf{T}})\ensuremath{\boldsymbol{\mathrm{\varphi}}}^{\mathrm{a}}&=-\ensuremath{\mathbf{K}}_{\Gamma_{\mathrm{Dir}}}\ensuremath{\boldsymbol{\mathrm{\varphi}}}^{\mathrm{Dir}},\label{eq:electric_deterministic_boundary}\\
        \ensuremath{ \mathbf{M} }_{\rho c}\dot{\ensuremath{\mathbf{T}}}+\left(\widehat{\ensuremath{\mathbf{K}}}_{\lambda}(\ensuremath{\mathbf{T}})+\ensuremath{ \mathbf{M} }^{\partial D}\right)\ensuremath{\mathbf{T}}&=\widehat{\mathbf{Q}}(\ensuremath{\boldsymbol{\mathrm{\varphi}}},\ensuremath{\mathbf{T}})+\ensuremath{ \mathbf{M} }^{\partial D}\ensuremath{\mathbf{T}}_{\infty}.
        \label{eq:thermal_deterministic_boundary}
    \end{align}
    \label{eq:electrothermal_deterministic_boundary}
\end{subequations}

\section{Relation to Flow Problems with Fissures}
\label{sec:error}
Let us focus on the 1D-3D coupling and consider a simplified elliptic model problem, which represents, for instance, the electrical subproblem.
We show that this model problem is closely related to a 1D-3D coupled problem of blood flow through tissues with thin tubular structures for the vessels as analyzed in~\cite{DAngelo_2008aa,DAngelo_2012aa}.
More precisely, we show that the simplified elliptic wire problem is obtained in the limit of infinite permeability of the vessel-tissue interface.
We thereby rely on the well-known similarities between the \gls{FE} and the present \gls{FIT} approach~\cite{Bossavit_2000ab}.
Several papers consider a singular Dirac\xspace right hand side for elliptic problems in an \gls{FE} context~\cite{Apel_2011aa,Babuska_1969aa}.
It should be emphasized that in the present approach, as well as in~\cite{DAngelo_2008aa,DAngelo_2012aa}, the singular 1D contribution depends on the solution itself, which further complicates the problem.

In a first step, we introduce a weak formulation together with the associated solution spaces for the model problem, where, for simplicity, we assume a linear conductivity.
Furthermore, we consider only one wire $\Lambda=\{\mathbf{x}(s),s\in\overline{\Lambda}=(0,1)\}$, such that its index can be omitted.
We denote with $u$ and $\overline{u}$ the 3D and 1D component of
\begin{subequations}
    \begin{alignat}{2}
        -\nabla \cdot\left(\alpha\nabla u\right)&=q_{\alpha}(\overline{u})\delta_{\Lambda}+f,\label{eq:elliptic_model3D}\\
        -\frac{\mathrm{d}}{\mathrm{d}s}\left(\overline{\alpha}\frac{\mathrm{d}}{\mathrm{d}s}\overline{u}\right)&=\overline{q}_{\alpha}(u).\label{eq:elliptic_model1D}
    \end{alignat}
    \label{eq:elliptic_model}\end{subequations}
A major difficulty in the analysis is the singularity of $u$ at $\Lambda$.
In particular, $u$ exhibits a logarithmic singularity and does not belong to the standard space $H^{1}(D)$, see Section~\ref{subsec:couplingOperator}.
Instead, the solution space can be identified with the weighted Sobolev\xspace space $V_{\delta}=H^{1}_{\delta}(D)$, see~\cite[Section 2]{DAngelo_2012aa}, where the weight is given by the distance to $\Lambda$ to the power of $2\delta$ with $\delta\in(0,1)$.
On the other hand, the 1D-variables belong to the space $\overline{V}:=H^{1}(\overline{\Lambda})$.
We consider homogeneous Dirichlet\xspace boundary conditions on $D$ for simplicity.
The weak form of \eqref{eq:elliptic_model3D} reads, find $u\in V_{\delta}$, such that
\begin{equation}
    \langle\alpha\nabla u,\nabla v\rangle_{D}=(q_{\alpha},\text{tr}_{\Lambda}v)_{\Lambda}+(f,v)_{D},\quad\forall v\in V_{-\delta},
    \label{eq:elliptic_3d_weak}
\end{equation}
where $\langle\cdot,\cdot\rangle_{D}$ refers to the duality product of $V_{\delta}$ and $V_{-\delta}$, whereas $(\cdot,\cdot)_{\{\Lambda,D\}}$ refers to the $L^{2}$-inner product on $\{\Lambda,D\}$.
Also, $\text{tr}_{\Lambda}$ denotes the trace operator from $V_{-\delta}$ to $L^{2}(\Lambda)$.
In the setting of this paper, the left hand side of \eqref{eq:elliptic_3d_weak} represents the divergence of the 3D current (heat flow), whereas the first term on the right hand side represents the current (flow) transport density from the 1D subdomain into the 3D surrounding.
The weak form of \eqref{eq:elliptic_model1D} reads, find $\overline{u}\in\overline{V}$, subject to
\begin{equation}
    \left(\overline{\alpha}\frac{\mathrm{d}}{\mathrm{d}s}\overline{u},\frac{\mathrm{d}}{\mathrm{d}s}\overline{v}\right)_{\overline{\Lambda}}=(\overline{q}_{\alpha},\overline{v})_{\overline{\Lambda}},\quad\forall\overline{v}\in\overline{V},
    \label{eq:elliptic_1d_weak}
\end{equation}
see~\cite{DAngelo_2012aa} for details.
In~\cite{DAngelo_2008aa}, the law for $\overline{q}_{\alpha}$ is chosen as $\overline{q}_{\alpha}=\overline{\beta}((\Pi u)\circ\mathbf{x}-\overline{u})$, where $\overline{\beta}=|\frac{\mathrm{d}}{\mathrm{d}s}\mathbf{x}|\beta\circ\mathbf{x}$ and $\beta$ refers to the permeability coefficient for the vessel-tissue blood transfer.
Note that the densities $q_{\alpha}$ and $\overline{q}_{\alpha}$ in \eqref{eq:elliptic_3d_weak} and \eqref{eq:elliptic_1d_weak} are related through $\overline{q}_{\alpha}=|\frac{\mathrm{d}}{\mathrm{d}s}\mathbf{x}|q_{\alpha}\circ\mathbf{x}$.

The coupled formulation reads, find $(u,\overline{u})\in V_{\delta}\times\overline{V}$ subject to
\begin{equation}
    \langle\alpha\nabla u,\nabla v\rangle_{D}+\left(\overline{\alpha}\frac{\mathrm{d}}{\mathrm{d}s}\overline{u},\frac{\mathrm{d}}{\mathrm{d}s}\overline{v}\right)_{\Lambda}=\left(\overline{\beta}(\mathbf{x}^{*}\Pi u-\overline{u}),\mathbf{x}^{*}v-\overline{v}\right)_{\Lambda}+(f,v)_{D},
    \label{eq:variational}
\end{equation}
for all $(v,\overline{v})\in V_{-\delta}\times\overline{V}$.
The corresponding \gls{FE} formulation reads, find $(u_{h},\overline{u}_{h})\in V_{h}\times\overline{V}_{h}$, subject to 
\begin{equation}
    \langle\alpha\nabla u_{h},\nabla v_{h}\rangle_{D}+\left(\overline{\alpha}\frac{\mathrm{d}}{\mathrm{d}s}\overline{u}_{h},\frac{\mathrm{d}}{\mathrm{d}s}\overline{v}_{h}\right)_{\overline{\Lambda}}
    =\left(\overline{\beta}(\mathbf{x}^{*}\Pi u_{h}-\overline{u}_{h}),\mathbf{x}^{*}v_{h}-\overline{v}_{h}\right)_{\overline{\Lambda}}+(f,v_{h})_{D},
    \label{eq:variational_fem}
\end{equation}
for all $(v_{h},\overline{v}_{h})\in V_{h}\times\overline{V}_{h}$, representing standard piecewise linear, globally continuous polynomial basis functions.
The 3D polynomials coincide with the Whitney\xspace basis functions of Section~\ref{sec:continuous_system}.
The connection from \gls{FE} to \gls{FIT} is established by introducing grid dependent inner products as follows 
\begin{align}
    \langle\alpha u_{h},v_{h}\rangle_{D,h}&:=\mathbf{v}^{\top}\ensuremath{ \mathbf{M} }_{\alpha}\mathbf{u},\\
    (\overline{\alpha}\,\overline{u}_{h},\overline{v}_{h})_{\overline{\Lambda},h}&=\overline{\mathbf{v}}^{\top}\overline{\ensuremath{ \mathbf{M} }}_{\alpha}\overline{\mathbf{u}}=\overline{\mathbf{v}}^{\top}\text{diag}\left(\frac{\overline{\alpha}_{1}}{|\Lambda_{1}|},\ldots,\frac{\overline{\alpha}_{N^{\mathrm{1D}}-1}}{|\Lambda_{N^{\mathrm{1D}}-1}|}\right)\overline{\mathbf{u}},\\ 
    (\overline{\beta}\,\overline{u}_{h},\overline{v}_{h})_{\overline{\Lambda},h}&=\overline{\mathbf{v}}^{\top}\overline{\ensuremath{ \mathbf{M} }}_{\beta}\overline{\mathbf{u}}=\overline{\mathbf{v}}^{\top}\text{diag}\left(\overline{\beta}_{1}|\protect\widetilde{\Lambda}_{1}|,\ldots,\overline{\beta}_{N^{\mathrm{1D}}}|\protect\widetilde{\Lambda}_{N^{\mathrm{1D}}}|\right)\overline{\mathbf{u}}.
\end{align}
Here, $\protect\widetilde{\Lambda}_{j}$ is the 1D dual element given by half of $\Lambda_{j-1}$ and $\Lambda_{j}$.
This \gls{FE}-\gls{FIT} connection can be viewed as applying the trapezoidal rule in each dimension, see~\cite{Chung_2005aa}\footnote{For a detailed explanation, see section 6 in the preprint of~\cite{Chung_2005aa}.} for details.
Then, we seek $(u_{h}^{\FIT},\overline{u}_{h}^{\FIT})\in V_{h}\times\overline{V}_{h}$, subject to 
\begin{multline}
    \langle\alpha\nabla u_{h}^{\FIT},\nabla v_{h}\rangle_{D,h}+\left(\overline{\alpha}\frac{\mathrm{d}}{\mathrm{d}s}\overline{u}_{h}^{\FIT},\frac{\mathrm{d}}{\mathrm{d}s}\overline{v}_{h}\right)_{\overline{\Lambda},h}=\\
    \left(\overline{\beta}(\mathbf{x}^{*}\Pi u_{h}^{\FIT}-\overline{u}_{h}^{\FIT}),\mathbf{x}^{*}v_{h}-\overline{v}_{h}\right)_{\overline{\Lambda},h}+(f,v_{h})_{D,h}.
    \label{eq:variational_fit}
\end{multline}
Problem \eqref{eq:variational_fit} is equivalent to the matrix system of equations
\begin{subequations}
    \begin{align}
        \G^{\top}\ensuremath{ \mathbf{M} }_{\alpha}\G\mathbf{u}&=\ensuremath{ \mathbf{R} }_{\text{N}}^{\top}\overline{\ensuremath{ \mathbf{M} }}_{\beta}\left(\boldsymbol{\Pi}\mathbf{u}-\overline{\mathbf{u}}\right)+\mathbf{f},\label{eq:coupled_system_FIT_3D}\\
        \overline{\ensuremath{\mathbf{P}}}_{s}^{\top}\overline{\ensuremath{ \mathbf{M} }}_{\alpha}\overline{\ensuremath{\mathbf{P}}}_{s}\overline{\mathbf{u}}&=-\overline{\ensuremath{ \mathbf{M} }}_{\beta}\left(\boldsymbol{\Pi}\mathbf{u}-\overline{\mathbf{u}}\right).\label{eq:coupled_system_FIT_1D} 
    \end{align}
\label{eq:coupled_system_FIT}\end{subequations}
Multiplying \eqref{eq:coupled_system_FIT_1D} with $\ensuremath{ \mathbf{R} }_{\text{N}}^{\top}$ and adding it to \eqref{eq:coupled_system_FIT_3D} we obtain
\begin{equation}
    \G^{\top}\ensuremath{ \mathbf{M} }_{\alpha}\G\mathbf{u}+\ensuremath{ \mathbf{R} }_{\text{N}}^{\top}\overline{\ensuremath{\mathbf{P}}}_{s}^{\top}\overline{\ensuremath{ \mathbf{M} }}_{\alpha}\overline{\ensuremath{\mathbf{P}}}_{s}\overline{\mathbf{u}}=\mathbf{f},
    \label{eq:coupled_system_FIT_combined} 
\end{equation}
which is the same as \eqref{eq:dae_detailed} without the transient part and the source term $\mathbf{f}$.
It should be noted at this point that our coupling does not require the implementation of the complicated boundary term on the right hand side of~\eqref{eq:variational_fem}.
Contrary to \eqref{eq:dae_detailed}, the coupling condition \eqref{eq:discrete_projection} is not directly contained in \eqref{eq:coupled_system_FIT}, but recovered in the limit $\beta\rightarrow\infty$.
In our setting, $\beta$ represents a contact conductivity and hence, we consider the limit of perfect conduction between 1D and 3D part.
Indeed, \eqref{eq:coupled_system_FIT_1D} is equivalent to 
\begin{equation}
    -\left(\overline{\ensuremath{ \mathbf{M} }}_{\beta}\right)^{-1}\overline{\ensuremath{\mathbf{P}}}_{s}^{\top}\overline{\ensuremath{ \mathbf{M} }}_{\alpha}\overline{\ensuremath{\mathbf{P}}}_{s}\overline{\mathbf{u}}=\left(\boldsymbol{\Pi}\mathbf{u}-\overline{\mathbf{u}}\right)
    \label{eq:coupled_system_coupling}
\end{equation}
and passing to the limit $\beta\rightarrow\infty$ yields $\boldsymbol{\Pi}\mathbf{u}-\overline{\mathbf{u}}=\ensuremath{\mathbf{0}}.$

A strategy to bound the total error of the numerical scheme could consist in using the relation to the \gls{FE} method, established above, and the triangle inequality as
\begin{equation}
    \|u-u_{h}^{\FIT}\|\leq\|u-u_{h}\|+\|u_{h}-u_{h}^{\FIT}\|. 
    \label{eq:FITerror}
\end{equation}
D'Angelo~\cite{DAngelo_2012aa} has shown that for the \gls{FE} error of the 3D variable there holds
\begin{equation}
    \|u -u_{h} \| \leq C_{1} h \|u \|_{V^2_{1+\varepsilon}} + C_{2} h \|\hat{u} \|_{H^{1}(\Lambda)}, 
    \label{eq:dangelo}
\end{equation}
with $\varepsilon \in (0, \delta)$, provided that the solution is sufficiently regular, $\beta$ small enough and the mesh grading strong enough.
The same estimate was established for the 1D variable in a suitable norm.
The norm $\| \cdot \|$ appearing in (\ref{eq:dangelo}) is defined as 
\begin{equation}
    \|\cdot \| := \|\alpha^{1/2} \nabla \cdot\|_{L^2_\delta(D)},
\end{equation}
where in $L^{2}_{\delta}(D)$ the weight is again given by the distance to $\overline{\Lambda}$ to the power of $2\delta$.
Moreover, $V^{2}_{1+\varepsilon}$ refers to a Kondratiev\xspace-type weighted space, see~\cite{DAngelo_2012aa} for details.
The second term on the right hand side of~\eqref{eq:FITerror}, representing the difference between \gls{FE} method and \gls{FIT}, is of order $\mathcal{O}(h)$, which can be obtained by bounding the error associated to the trapezoidal rule.
Yet, it remains to show that the system \eqref{eq:coupled_system_FIT_combined}, \eqref{eq:coupled_system_coupling} is well-posed.
Additionally, the limit $\beta\rightarrow\infty$ for the continuous problem \eqref{eq:variational} and its \gls{FE} approximation~\eqref{eq:variational_fem} is not included in~\cite{DAngelo_2012aa} and has to be analyzed separately, which is beyond the scope of this paper. 

 \section{Numerical Examples and Application}
\label{sec:numerics}

In this section, the proposed method is validated using stationary model problems and a transient electrothermal microelectronic chip package.
For all implementations, the temperature dependence of the materials is neglected and thus, linear problems are considered.
Before presenting the results for the individual models, we define local and global gradings of the grid.
For all examples used here, the Matlab\textsuperscript{\textregistered}\xspace code to generate the presented results is openly available~\cite{Casper_2018ae}.

\subsection{Grid Generation with Local and Global Grading}
\label{sec:gridGeneration}

As introduced in Section~\ref{sec:discrete_system}, we apply the \gls{FIT} on a pair of rectilinear 3D grids.
Any 3D grid that fulfills these properties is denoted by $\pazocal{G}$ and is constructed using a Cartesian\xspace product of 1D grids.
For all grids $\pazocal{G}$, the average 3D edge length is denoted by $h$.
On the other hand, a 1D grid is used for the discretization of the wires and is denoted by $\pazocal{\overline{G}}$.
We apply the same 1D grid for all wires with respect to their parametrizations.
The 1D grid $\overline{\pazocal{G}}$ is chosen to be equidistant in terms of the wires' parametrization $s$ with the step size denoted by $\overline{h}$.
Note that for the numerical implementation, the 1D grid points of $\overline{\pazocal{G}}$ coincide with 3D grid points of $\pazocal{G}$.
Therefore, for straight wires and equidistant 3D grids, the 1D step size $\overline{h}$ is always a multiple of the 3D step size $h$.

\begin{figure}[b!]
    \centering
    \begin{subfigure}{0.325\textwidth}
        \centering
        \includegraphics{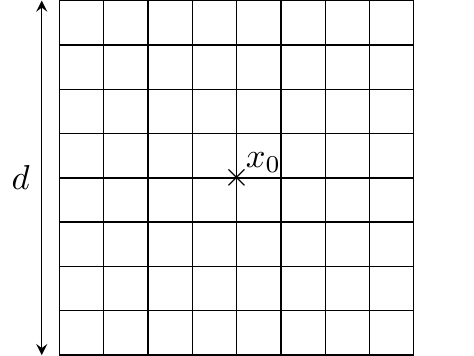}
        \caption{}
        \label{fig:gridEquidistant}
    \end{subfigure}
    \begin{subfigure}{0.325\textwidth}
        \centering
        \includegraphics{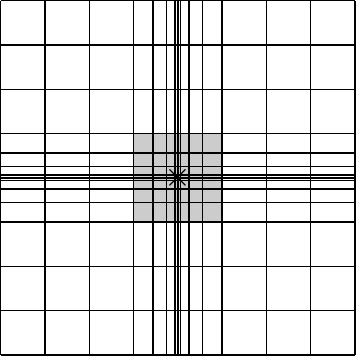}
        \caption{}
        \label{fig:gridLocalGrading}
    \end{subfigure}
    \begin{subfigure}{0.325\textwidth}
        \centering
        \includegraphics{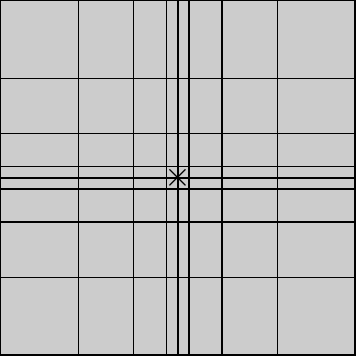}
        \caption{}
        \label{fig:gridGlobalGrading}
    \end{subfigure}
    \caption{Cross section view of (a) an equidistant 3D grid $\pazocal{G}_{1}$, (b) a locally graded grid $\pazocal{G}_{0.5,4}^{d/8}$ and (c) a globally graded grid $\pazocal{G}_{0.5}$, where the region of refinement $D_{\text{r}}$ is shown in gray.}
\end{figure}

Due to the singular solution of \eqref{eq:electrothermal}, a graded grid is required to recover the expected convergence rates.
Let us consider a 3D domain $D$ with a line on which the solution becomes singular.
We assume that an initial, equidistant grid $\pazocal{G}_{1}$ is given, see Figure~\ref{fig:gridEquidistant}.
The region on which the refinement is applied is denoted by $D_{\text{r}}\subset D$.
As the 3D grid is composed by the Cartesian product of individual 1D grids, the refinement strategy for a 1D grid is described here.
Based on \cite{Apel_1996aa}, the 1D refinement around a singular point $x_{0}$ is applied using the layers
\begin{equation}
    r_{i}=b\left(\frac{i}{\pazocal{N}}\right)^{\frac{1}{\mu}},\quad\text{with}\quad i=0,\dots,\pazocal{N},
    \label{eq:localRefinement}
\end{equation}
where $\pazocal{N}$ is the number of refinement layers, $\mu$ the grading and $b$ the radius of the refinement.
Note that $\mu=1$ results in an equidistant refinement whereas $\mu\rightarrow 0$ gives a stronger grading towards the singularity.
Furthermore, $b$ is chosen such that no grid points of $\pazocal{G}_{1}$ fall in the refinement region $D_{\text{r}}$.
The only exception to this rule is a possible grid point at $x_{0}$.
Note that for a refinement with $\pazocal{N}$ refinement layers with the assumption that $x_{0}$ was part of $\pazocal{G}_{1}$, $2\pazocal{N}$ 1D points are added.
Due to the rectilinear 3D grid, this results in a propagation of the refinement along the coordinate directions.
Thus, $2\pazocal{N}$ additional 1D points result in $(2\pazocal{N}+1)^{2}-1$ additional points for a two-dimensional refinement and $(2\pazocal{N}+1)^{3}-1$ additional points for a three-dimensional refinement.
For line sources that are the main subject of this paper, a two-dimensional refinement is required.
After applying the tensor product on the individually refined 1D grids, a locally graded 3D grid is obtained and denoted by $\pazocal{G}_{\mu,\pazocal{N}}^{b}$, see Figure~\ref{fig:gridLocalGrading}. 

With the local grading as introduced above, a global grid grading can be defined as a specific choice of the local grading.
Let an initial grid contain the singular point only.
Then, by choosing the refinement radius $b=d/2$, where $d$ is the width of $D$, the refinement given by \eqref{eq:localRefinement} results in a refinement on the entire computational domain $D$ such that $D_{\text{r}}=D$.
A grid obtained by this special choice is denoted by $\pazocal{G}_{\mu}$, see Figure~\ref{fig:gridGlobalGrading}.
Note that an additional choice of $\mu=\num{1}$ results in an equidistant grid $\pazocal{G}_{1}$ without any refinement nor grading and the same number of grid points as $\pazocal{G}_{\mu}$.

\subsection{Numerical Examples}
\label{subsec:numericalExamples}

The model problems consist of one wire embedded in a homogeneous cube such that an exact representation of the geometry using a rectilinear grid is possible.
In particular, three different models are considered.
First, the coupling is realized point-wise using an external circuitry resulting in a 0D-2D coupling approach. 
Secondly, a straight wire embedded in the cube is simulated in a 1D-3D coupling approach.
Thirdly, the general case of a bent wire is considered.
For the first two models, an analytical reference solution is available while a fine reference is used to study the convergence of the third model.
For these model problems, the electrical problem given by \eqref{eq:electric_deterministic_boundary} is solved.

The considered cube $D$ is of side length $d=\SI{1}{m}$ and conductivity $\sigma=\SI{1}{S\per m}$.
In all examples, the wire's radius is $\overline{r}=\SI{1}{\mu m}$ and its cross-sectional area is $\overline{A}=\pi\overline{r}^2$.
The conductivity of the wire is given by $\overline{\sigma}=\num{1e15}\overline{A}\sigma$ manifesting an electrical example opposing thermal examples for which $\overline{\sigma}\approx\num{1e2}\overline{A}\sigma$.
As discussed in Section~\ref{subsec:couplingOperator}, the coupling coefficient $\gamma$ is used with a reference radius $r_{0}$ that we choose as the radius of an equivalent\footnote{The term equivalent refers to a geometry of equal cross-sectional area as the cube} cylindrical (or circular in the 2D case) geometry such that $r_{0}=\sqrt{d^{2}/\pi}$.
As a reference, Table~\ref{tab:simSettings} summarizes the described parameters.

\begin{table}[t]
    \centering
    \begin{tabular}{ccc}\toprule
        Parameter           & Description                               & Value\\\bottomrule\toprule
        $d$                 & Width of $D$                         & \SI{1}{m}\\
        $\sigma$            & Conductivity of $D$                 & \SI{1}{S\per m}\\
        $\overline{r}$            & Radius of wire                            & \SI{1}{\mu m}\\
        $\overline{A}$            & Cross-sectional area of wire              & $\pi\overline{r}^{2}$\\
        $\overline{\sigma}$        & Conductivity of $\Lambda$                 & $\num{1e15}\overline{A}\sigma$\\
        $r_{0}$             & Reference radius                          & $\sqrt{d^{2}/\pi}$\\
        \bottomrule
    \end{tabular}
    \caption{Parameters with description and value as used for the numerical examples.}
    \label{tab:simSettings}
\end{table}

To quantify the errors of the method, we introduce the following error measures.
Let $\ensuremath{\boldsymbol{\mathrm{\varphi}}}_{h}$ and $\overline{\ensuremath{\boldsymbol{\mathrm{\varphi}}}}_{h}$ denote the 3D and 1D \gls{FIT} solution vectors, respectively, and $\varphi$ and $\overline{\varphi}$ the corresponding analytical solutions.
Since we neglect the Joule\xspace losses in the 3D domain (cf. Section~\ref{sec:continuous_system}), we consider the error in the solution's derivative only for the 1D solution.
First, we define
\begin{equation}
    \ensuremath{\varepsilon_{L^{2}}^{\mathrm{1D}}}\xspace:=\frac{\lVert\overline{\ensuremath{\boldsymbol{\mathrm{\varphi}}}}_{h}-\overline{\ensuremath{\boldsymbol{\mathrm{\varphi}}}}\rVert_{L^{2},h}^{\mathrm{1D}}}{\lVert\overline{\ensuremath{\boldsymbol{\mathrm{\varphi}}}}\rVert_{L^{2},h}^{\mathrm{1D}}},
    \quad
    \ensuremath{\varepsilon_{H^{1}}^{\mathrm{1D}}}\xspace:=\frac{\lVert\ensuremath{\mathbf{P}}_{s}\overline{\ensuremath{\boldsymbol{\mathrm{\varphi}}}}_{h}-\ensuremath{\mathbf{P}}_{s}\overline{\ensuremath{\boldsymbol{\mathrm{\varphi}}}}\rVert_{L^{2},h}^{\mathrm{1D}}}{\lVert\ensuremath{\mathbf{P}}_{s}\overline{\ensuremath{\boldsymbol{\mathrm{\varphi}}}}\rVert_{L^{2},h}^{\mathrm{1D}}},
    \quad
    \ensuremath{\varepsilon_{L^{2}}^{\mathrm{3D}}}\xspace:=\frac{\lVert\ensuremath{\boldsymbol{\mathrm{\varphi}}}_{h}-\ensuremath{\boldsymbol{\mathrm{\varphi}}}\rVert_{L^{2},h}^{\mathrm{3D}}}{\lVert\ensuremath{\boldsymbol{\mathrm{\varphi}}}\rVert_{L^{2},h}^{\mathrm{3D}}},
    \label{eq:errorEps}
\end{equation}
where $\ensuremath{\boldsymbol{\mathrm{\varphi}}}$ and $\overline{\ensuremath{\boldsymbol{\mathrm{\varphi}}}}$ refer to the analytical solutions evaluated on the same grid as $\ensuremath{\boldsymbol{\mathrm{\varphi}}}_{h}$ and $\overline{\ensuremath{\boldsymbol{\mathrm{\varphi}}}}_{h}$.
Secondly,
\begin{subequations}
    \begin{align}
         \ensuremath{\delta_{L^{2}}^{\mathrm{1D}}}\xspace&:=\frac{\left|\lVert\overline{\ensuremath{\boldsymbol{\mathrm{\varphi}}}}_{h}\rVert_{L^{2},h}^{\mathrm{1D}}-\lVert\overline{\varphi}\rVert_{L^{2}}^{\mathrm{1D}}\right|}{\lVert\overline{\varphi}\rVert_{L^{2}}^{\mathrm{1D}}},
         \quad
         \ensuremath{\delta_{H^{1}}^{\mathrm{1D}}}\xspace:=\frac{\left|\lVert\ensuremath{\mathbf{P}}_{s}\overline{\ensuremath{\boldsymbol{\mathrm{\varphi}}}}_{h}\rVert_{L^{2},h}^{\mathrm{1D}}-\lVert\partial_{s}\overline{\varphi}\rVert_{L^{2}}^{\mathrm{1D}}\right|}{\lVert\partial_{s}\overline{\varphi}\rVert_{L^{2}}^{\mathrm{1D}}},\\
         \ensuremath{\delta_{L^{2}}^{\mathrm{3D}}}\xspace&:=\frac{\left|\lVert\ensuremath{\boldsymbol{\mathrm{\varphi}}}_{h}\rVert_{L^{2},h}^{\mathrm{3D}}-\lVert\varphi\rVert_{L^{2}}^{\mathrm{3D}}\right|}{\lVert\varphi\rVert_{L^{2}}^{\mathrm{3D}}},
    \end{align}
    \label{eq:errordelta}%
\end{subequations}
where the norm of the analytical solution is compared to the norm of the \gls{FIT} solution.
Lastly, if no analytical solution is available, we use
\begin{subequations}
    \begin{align}
         \ensuremath{\Delta_{L^{2}}^{\mathrm{1D}}}\xspace&:=\frac{\left|\lVert\overline{\ensuremath{\boldsymbol{\mathrm{\varphi}}}}_{h}\rVert_{L^{2},h}^{\mathrm{1D}}-\lVert\overline{\ensuremath{\boldsymbol{\mathrm{\varphi}}}}\rVert_{L^{2},h}^{\mathrm{1D}}\right|}{\lVert\overline{\ensuremath{\boldsymbol{\mathrm{\varphi}}}}\rVert_{L^{2},h}^{\mathrm{1D}}},
         \quad
         \ensuremath{\Delta_{H^{1}}^{\mathrm{1D}}}\xspace:=\frac{\left|\lVert\ensuremath{\mathbf{P}}_{s}\overline{\ensuremath{\boldsymbol{\mathrm{\varphi}}}}_{h}\rVert_{L^{2},h}^{\mathrm{1D}}-\lVert\ensuremath{\mathbf{P}}_{s}\overline{\ensuremath{\boldsymbol{\mathrm{\varphi}}}}\rVert_{L^{2},h}^{\mathrm{1D}}\right|}{\lVert\ensuremath{\mathbf{P}}_{s}\overline{\ensuremath{\boldsymbol{\mathrm{\varphi}}}}\rVert_{L^{2},h}^{\mathrm{1D}}},\\
         \quad
         \ensuremath{\Delta_{L^{2}}^{\mathrm{3D}}}\xspace&:=\frac{\left|\lVert\ensuremath{\boldsymbol{\mathrm{\varphi}}}_{h}\rVert_{L^{2},h}^{\mathrm{3D}}-\lVert\ensuremath{\boldsymbol{\mathrm{\varphi}}}\rVert_{L^{2},h}^{\mathrm{3D}}\right|}{\lVert\ensuremath{\boldsymbol{\mathrm{\varphi}}}\rVert_{L^{2},h}^{\mathrm{3D}}},
    \end{align}
    \label{eq:errorDelta}%
\end{subequations}
where $\ensuremath{\boldsymbol{\mathrm{\varphi}}}$ and $\overline{\ensuremath{\boldsymbol{\mathrm{\varphi}}}}$ are \gls{FIT} solutions computed on a very fine grid.
The discrete and continuous norms used in \eqref{eq:errorEps}--\eqref{eq:errorDelta} are defined by
\begin{alignat}{5}
      \lVert            \overline{\mathbf{u}}\rVert_{L^{2},h}^{\mathrm{1D}}   &:=      \sqrt{\overline{\mathbf{u}}^{\top}\overline{\ensuremath{ \mathbf{D} }}_{\protect\tilde{\text{S}}}                                  \overline{\mathbf{u}}},
    \quad
    &&\lVert\ensuremath{\mathbf{P}}_{s}\overline{\mathbf{u}}\rVert_{L^{2},h}^{\mathrm{1D}}  &&:=      \sqrt{\overline{\mathbf{u}}^{\top}\overline{\ensuremath{\mathbf{P}}}_{s}^{\top}\overline{\ensuremath{ \mathbf{D} }}_{\text{S}}^{-1}\overline{\ensuremath{\mathbf{P}}}_{s}\overline{\mathbf{u}}},
    \quad
    &&\lVert                     \mathbf{u} \rVert_{L^{2},h}^{\mathrm{3D}}&&:=\left(\sqrt{         \mathbf{u} ^{\top}\ensuremath{ \mathbf{D} }_{\protect\tilde{\text{V}}}                                               \mathbf{u}}\right)_{\Omega},\\
    \lVert              \overline{  u}\rVert_{L^{2}}  ^{\mathrm{1D}}   &:=      \sqrt{\int_{\overline{\Lambda}}               \overline{u}^{2}       \mathrm{d} s},
    \quad
    &&\lVert\partial_{s}\overline{  u}\rVert_{L^{2}}  ^{\mathrm{1D}}  &&:=      \sqrt{\int_{\overline{\Lambda}}\left(\partial_{s}\overline{u}\right)^{2}\mathrm{d} s},
    \quad
    &&\lVert                       u \rVert_{L^{2}}  ^{\mathrm{3D}}&&:=      \sqrt{\int_{\Omega}                            u^{2}        \mathrm{d}\mathbf{x}},
\end{alignat}where $\ensuremath{ \mathbf{D} }_{\protect\tilde{\text{V}}}$, $\overline{\ensuremath{ \mathbf{D} }}_{\text{S}}$ and $\overline{\ensuremath{ \mathbf{D} }}_{\protect\tilde{\text{S}}}$  are diagonal matrices with the 3D dual volumes, the 1D primal lengths and the 1D dual lengths on the diagonal, respectively.
These matrices coincide with $\ensuremath{\mathbf{M}_{\rho c}}$, $\overline{\ensuremath{ \mathbf{M} }}_{\beta}$ and $\overline{\ensuremath{ \mathbf{M} }}_{\sigma}^{-1}$ for homogeneous materials of unit value.
Furthermore, $\Omega=[0,0.45d]\times[0,d]\times[0,d]$ refers to the domain in which the 3D errors are evaluated and the notation $\left(\,\cdot\,\right)_{\Omega}$ restricts the discrete vectors/matrices to the evaluation domain\footnote{Note that, if required, additional grid points to resolve the evaluation domain $\Omega$ are inserted}.

\subsubsection{0D-2D Coupling}

\begin{figure}[t]
	\centering
	\begin{subfigure}{0.32\textwidth}
        \centering
        \includegraphics{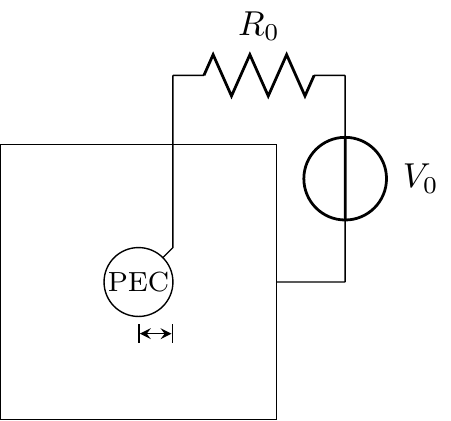}
        \caption{}
        \label{fig:resistorRectCpl2D}
    \end{subfigure}
	\begin{subfigure}{0.32\textwidth}
        \centering
        \includegraphics{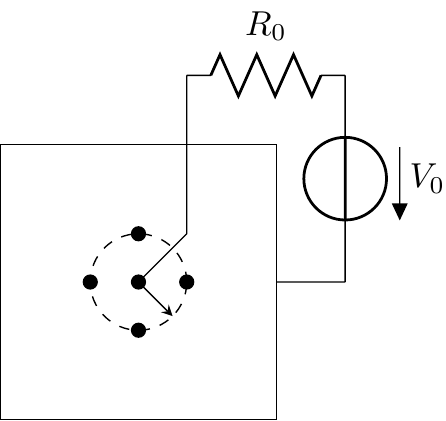}
        \caption{}
        \label{fig:resistorRectCpl2Dmodel}
    \end{subfigure}
	\begin{subfigure}{0.32\textwidth}
		\centering
		\includegraphics{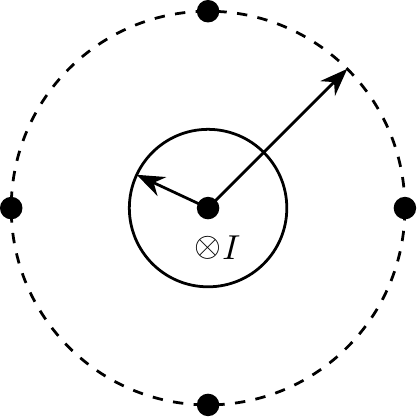}
        \caption{}
        \label{fig:wireCylAnnotated}
	\end{subfigure}
  	\caption{(a) shows the cross section of a brick-shaped resistor with wire-shaped inner PEC electrode connected to an external circuit. In (b), the inner electrode is replaced by a 0D representation and the coupling circle is shown as a dashed curve. (c) shows a magnification of the coupling circle while also depicting the actual wire radius $\overline{r}$ and the wire's current $I$.}
\end{figure}

As a first validation, we consider a brick-shaped resistor with parameters as given in Table~\ref{tab:simSettings}.
The \gls{PEC} wire of radius $\overline{r}$ and the surrounding boundary serve as the resistor's inner and outer electrodes.
The coupling is established by connecting the inner and outer electrode using a series connection of lumped resistor and voltage source, see Figure~\ref{fig:resistorRectCpl2D}.
Applying the parameters as shown in Table~\ref{tab:simSettings}, we use the thin wire assumption for the inner electrode and thus model it by a single point as shown in Figure~\ref{fig:resistorRectCpl2Dmodel}.
Then, the coupling is identified as a 0D-2D coupling.
The dashed circle depicts the coupling circle of radius $r_{\mathrm{cpl}}$ (cf. coupling condition \eqref{eq:cplCondition}) that is shown in a magnified view in Figure~\ref{fig:wireCylAnnotated}.
To avoid resolving the inner electrode, a coupling radius of $r_{\mathrm{cpl}}=\max_{l}\left(\ensuremath{L}_{l}^{xy}\right)\gg\overline{r}$ is chosen, where $\ensuremath{L}_{l}^{xy}$ iterates over the lengths of all edges perpendicular to the wire.

Assuming the resistance per unit length of the rectangular resistor to be given by
\begin{equation}
R_{\text{int}}'=\frac{\log\left(r_{0}/\overline{r}\right)}{2\pi\sigma},
\end{equation}
the resistance of the external resistor to be $R_{0}'=\SI{1}{\ohm m}$ and the applied voltage as $V_{0}=\SI{1}{V}$, the potential $\overline{\varphi}$ at the inner electrode is of interest.
For the current per unit length $I_{0}'=V_{0}/(R_{0}'+R_{\text{int}}')$ and a homogeneous conductivity $\sigma$, the analytical 2D solution of Laplace\xspace's equation is given by
\begin{equation}
    \varphi(r)=-\frac{I_{0}'}{2\pi\sigma}\log\left(\frac{r}{r_{0}}\right),
    \label{eq:0D2DrefSol}
\end{equation}
where $r_{0}=\sqrt{d^{2}/\pi}$ is the distance from the origin to the reference potential.
After applying the coupling condition \eqref{eq:cplCondition} to \eqref{eq:0D2DrefSol}, the 0D potential $\overline{\varphi}=\Pi\varphi$ is obtained.

\begin{figure}[t]
\centering
\begin{subfigure}{0.49\textwidth}
\centering
\includegraphics{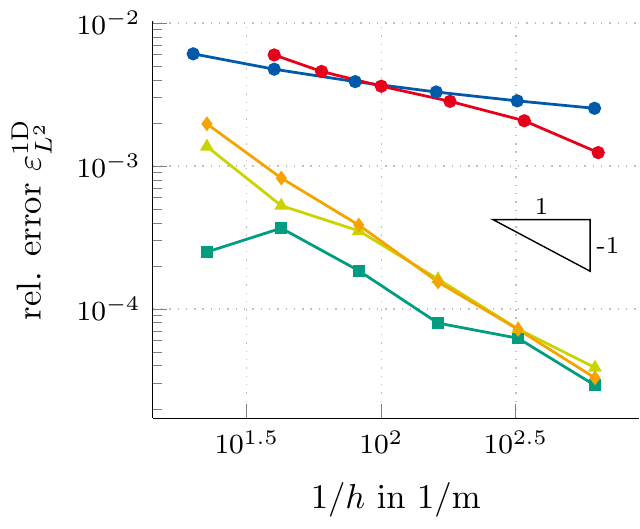}
\end{subfigure}
\begin{subfigure}{0.49\textwidth}
\centering
\includegraphics{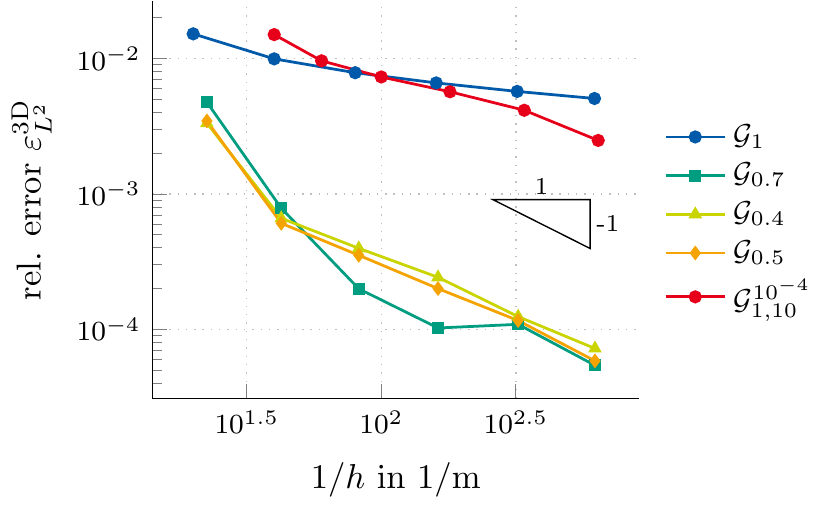}
\end{subfigure}
\caption{Convergence of (a) \ensuremath{\varepsilon_{L^{2}}^{\mathrm{1D}}}\xspace and (b) \ensuremath{\varepsilon_{L^{2}}^{\mathrm{3D}}}\xspace with respect to $h$ and different grid choices for the 0D-2D coupling.}
\label{fig:results2D}
\end{figure}

Applying \eqref{eq:ETdiscreteEl} to the here considered 0D-2D coupling, it simplifies to
\begin{equation}
    \ensuremath{\mathbf{G}}^{\top}\ensuremath{\mathbf{M}_{\sigma}}\ensuremath{\mathbf{G}}\ensuremath{\boldsymbol{\mathrm{\varphi}}}+\ensuremath{ \mathbf{R} }_{\text{N}}^{\top} G_{0}'\overline{\ensuremath{\boldsymbol{\mathrm{\varphi}}}}=\ensuremath{\mathbf{0}},
    \label{eq:coupled_system_FIT_2D}
\end{equation}
where, by abuse of notation, the matrices are 2D modifications of the usual 3D matrices and $G_{0}'=(R_{0}')^{-1}$.
We impose the reference solution $\varphi$ on the boundary of the domain (method of manufactured solution), apply the boundary conditions to \eqref{eq:coupled_system_FIT_2D} as described in Section~\ref{subsec:boundary} and solve the resulting system to obtain the \gls{FIT} solution $\overline{\ensuremath{\boldsymbol{\mathrm{\varphi}}}}_{h}$.
In this section, $h$ is the average edge length of all edges in $x$- and $y$-direction.
Using the analytical solution of \eqref{eq:0D2DrefSol} as reference, the convergence of the relative errors \ensuremath{\varepsilon_{L^{2}}^{\mathrm{3D}}}\xspace and \ensuremath{\varepsilon_{L^{2}}^{\mathrm{1D}}}\xspace for different grids with respect to $h$ is shown in Figure~\ref{fig:results2D}.
Considering the results for \ensuremath{\varepsilon_{L^{2}}^{\mathrm{1D}}}\xspace, we first observe that a uniform grid $\pazocal{G}_{1}$ gives a convergence rate far lower than one.
Secondly, if we use a globally graded grid, the convergence order can be improved to a value around one.
Thirdly, a locally graded grid $\pazocal{G}_{1,10}^{10^{-4}}$ gives only a slightly higher convergence rate than $\pazocal{G}_{1}$.
The convergence rates for the 2D error \ensuremath{\varepsilon_{L^{2}}^{\mathrm{3D}}}\xspace behave similarly.

\subsubsection{3D-1D Straight Wire Coupling}

\begin{figure}[t]
    \centering
    \begin{subfigure}{0.49\textwidth}
        \includegraphics{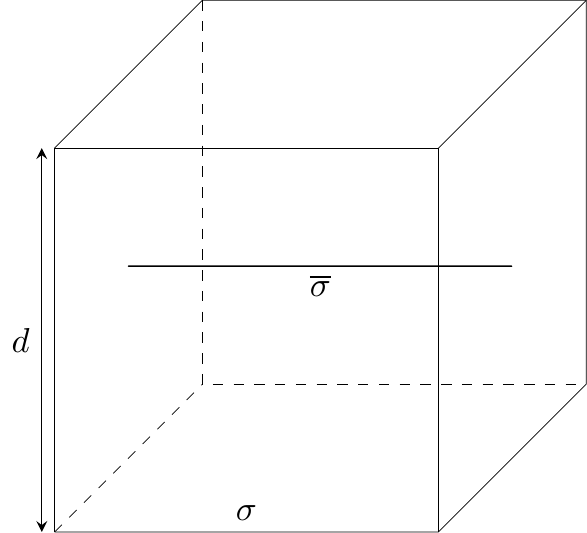}
        \caption{}
        \label{fig:wireStraight}
    \end{subfigure}
    \begin{subfigure}{0.49\textwidth}
        \includegraphics{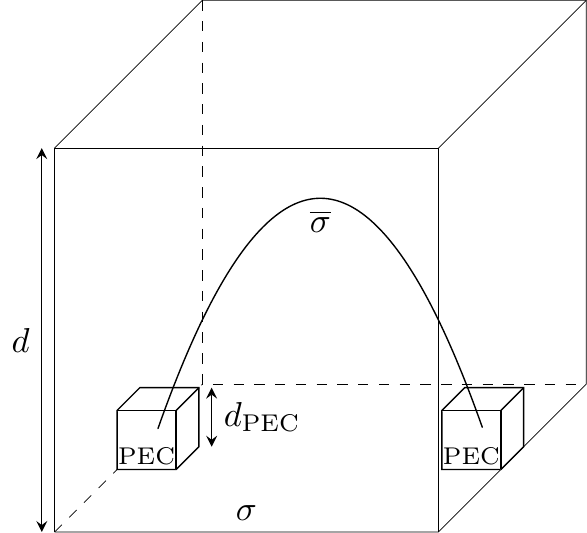}
        \caption{}
        \label{fig:wireBent}
    \end{subfigure}
    \caption{(a) Straight and (b) bent wire embedded in a cube of side length $d$. In both cases, the conductivities of cube and wire are annotated and in (b), additional \gls{PEC} cubes are used.}
    \label{fig:wireGeometries}
\end{figure}

In this section, we consider a 3D-1D coupling and apply the \gls{FIT} to solve the electrical problem.
The investigated model consists of a straight wire in $z$-direction positioned in the $xy$-center of a cube of side length $d$ that we refer to as the computational domain $D$, see Figure~\ref{fig:wireStraight}.
We again use the parameters as summarized in Table~\ref{tab:simSettings}.
Additionally, for the locally refined grids $\pazocal{G}_{1,10}^{b}$ and $\pazocal{G}_{0.5,10}^{b}$, we use $b=\frac{1}{3}\min_{l}\left(\ensuremath{L}_{l}^{xy}\right)$.  
For this setting, according to Section~\ref{subsec:couplingOperator}, admissible 3D and 1D solutions are given by
\begin{align}
    \varphi(r,z)=-\frac{I'(z)}{2\pi\sigma}\log\left(\frac{r}{r_{0}}\right)\quad\text{and}\quad\overline{\varphi}(z)=-\frac{I'(z)}{2\pi\sigma}\log\left(\frac{\overline{r}}{r_{0}}\right),
    \label{eq:1D3DrefSol}
\end{align}
where we choose $I'(z)=I_{0}'z/d$ and $I_{0}'=\SI{1}{A\per m}$.

\begin{figure}[p]
    \centering
    \begin{subfigure}{0.49\textwidth}
        \includegraphics{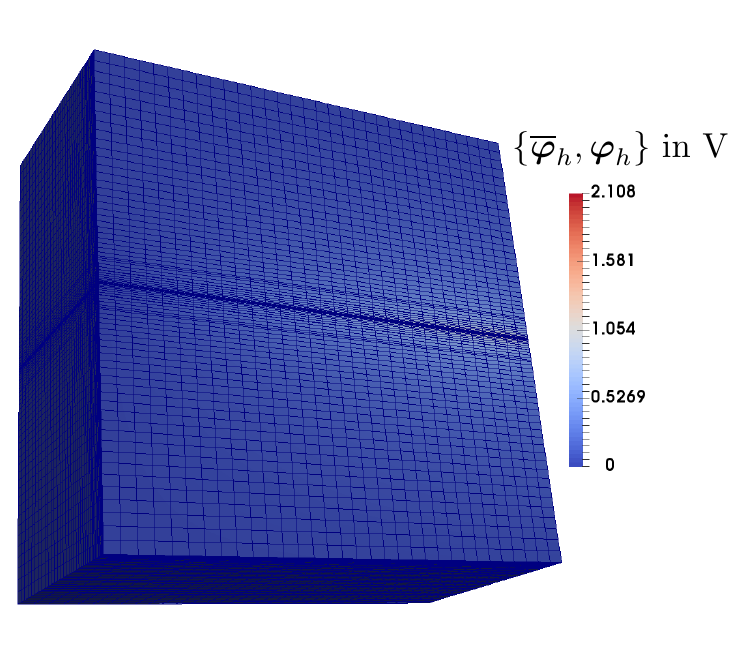}
                \caption{}
        \label{fig:resultsStraight3DplotMethod5}
    \end{subfigure}
    \begin{subfigure}{0.49\textwidth}
        \includegraphics{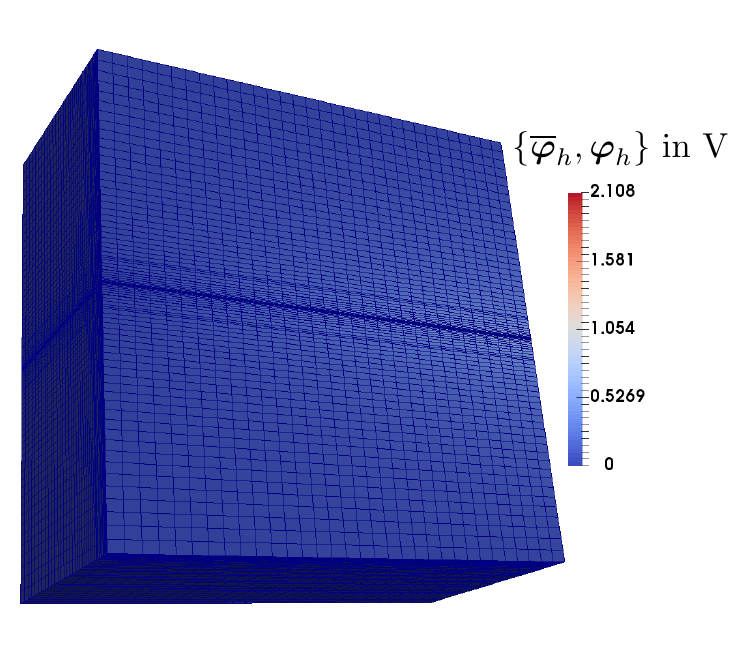}
                \caption{}
        \label{fig:resultsStraight3DplotMethod2}
    \end{subfigure}
    \caption{Values of $\overline{\ensuremath{\boldsymbol{\mathrm{\varphi}}}}_{h}$ on $\Lambda$ and $\ensuremath{\boldsymbol{\mathrm{\varphi}}}_{h}$ on $D\setminus\Lambda$ for $\pazocal{G}_{0.5}$, $h\approx\SI{1.822e-2}{m}$ and $\overline{h}=\num{3.125e-2}$ for (a) $r_{\mathrm{cpl}}=\SI{0}{m}$ and (b) $r_{\mathrm{cpl}}=\max_{l}(\ensuremath{L}_{l}^{xy})$.}
    \label{fig:resultsStraight3Dplot}
\end{figure}
\begin{figure}[p]
    \centering
    \begin{subfigure}{0.49\textwidth}
        \includegraphics{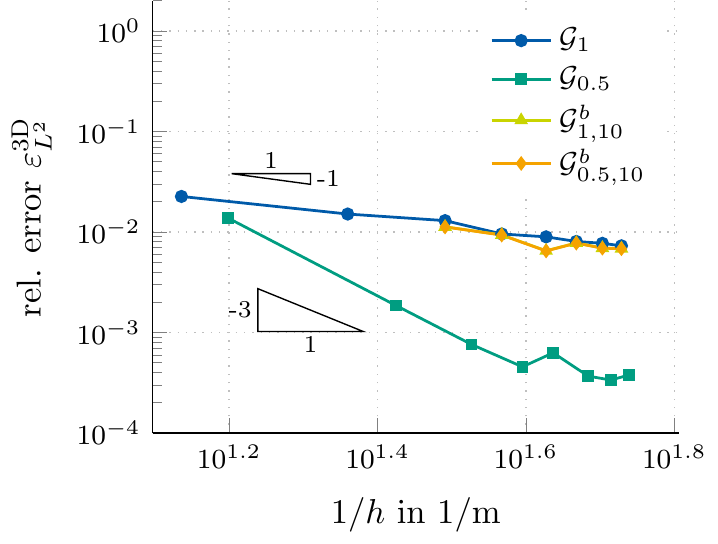}
                \caption{}
        \label{fig:resultsStraightepsL23Dgrid}
    \end{subfigure}
    \begin{subfigure}{0.49\textwidth}
        \includegraphics{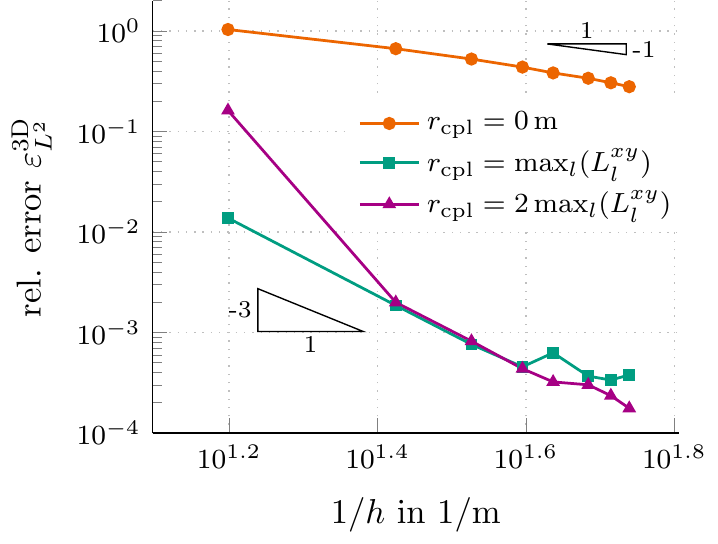}
                \caption{}
        \label{fig:resultsStraightepsL23DrCpl}
    \end{subfigure}
    \caption{Convergence of \ensuremath{\varepsilon_{L^{2}}^{\mathrm{3D}}}\xspace with respect to $h$ for $\overline{h}=\num{3.125e-2}$ and (a) different grid choices for the straight wire 1D-3D coupling and (b) for $\pazocal{G}_{0.5}$ and different choices of $r_{\mathrm{cpl}}$.}
    \label{fig:resultsStraightepsL23D}
\end{figure}

\begin{figure}[t]
    \centering
    \includegraphics{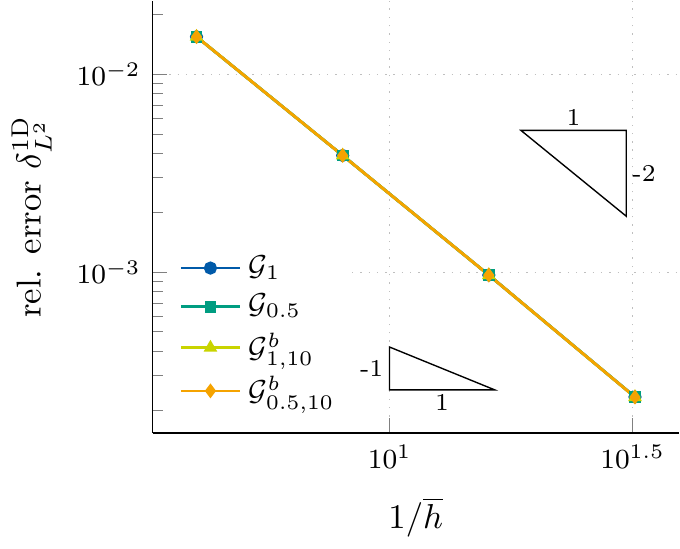}
        \caption{Convergence of \ensuremath{\varepsilon_{L^{2}}^{\mathrm{1D}}}\xspace with respect to $\overline{h}$ for a fixed but different $h$ for each 3D grid and for the straight wire 1D-3D coupling.}
    \label{fig:resultsStraightdeltaL21D}
\end{figure}

The analytical solution \eqref{eq:1D3DrefSol} is impressed on the boundary $\partial D$.
Then, \eqref{eq:electric_deterministic_boundary} is solved for the 3D solution $\ensuremath{\boldsymbol{\mathrm{\varphi}}}_{h}$ and the 1D solution $\overline{\ensuremath{\boldsymbol{\mathrm{\varphi}}}}_{h}$.
In Figure~\ref{fig:resultsStraight3Dplot}, $\overline{\ensuremath{\boldsymbol{\mathrm{\varphi}}}}_{h}$ is plotted on $\Lambda$ while $\ensuremath{\boldsymbol{\mathrm{\varphi}}}_{h}$ is plotted on $D\setminus\Lambda$ for $\pazocal{G}_{0.5}$, $h\approx\SI{1.822e-2}{m}$ and $\overline{h}=\num{3.125e-2}$ comparing $r_{\mathrm{cpl}}=\SI{0}{m}$ and $r_{\mathrm{cpl}}=\max_{l}(\ensuremath{L}_{l}^{xy})$.
The convergence of the relative errors as defined by \eqref{eq:errorEps} and \eqref{eq:errordelta} is investigated.
In Figures~\ref{fig:resultsStraightepsL23D} and \ref{fig:resultsStraightdeltaL21D}, the convergence of the errors \ensuremath{\varepsilon_{L^{2}}^{\mathrm{3D}}}\xspace and \ensuremath{\delta_{L^{2}}^{\mathrm{1D}}}\xspace is shown.
In Figure~\ref{fig:resultsStraightepsL23Dgrid}, the convergence of \ensuremath{\varepsilon_{L^{2}}^{\mathrm{3D}}}\xspace with respect to $h$ is shown for $\overline{h}=\num{3.125e-2}$, $r_{\mathrm{cpl}}=\max_{l}(\ensuremath{L}_{l}^{xy})$ and different choices of the 3D grid.
For the local gradings $\pazocal{G}_{1,10}^{b}$ and $\pazocal{G}_{0.5,10}^{b}$, the asymptotic convergence order is around one, independent of $\mu$ and similar to the order for a global grid grading $\pazocal{G}_{1}$.
For a globally graded grid $\pazocal{G}_{0.5}$, we observe a significantly higher convergence order than for an equidistant grid $\pazocal{G}_{1}$.
In Figure~\ref{fig:resultsStraightepsL23DrCpl}, the influence of the coupling radius $r_{\mathrm{cpl}}$ on the error and on the convergence order is investigated.
For a zero coupling radius, the 1D solution is taken directly from the 3D solution in the coupling points where the 3D solution is singular.
Therefore, as expected, the error is large and the convergence very slow.
For $r_{\mathrm{cpl}}>\SI{0}{m}$, a convergence order of around three is observed while the solution and the order is independent on the exact choice of $r_{\mathrm{cpl}}$.
In Figure~\ref{fig:resultsStraightdeltaL21D}, the convergence of \ensuremath{\delta_{L^{2}}^{\mathrm{1D}}}\xspace is shown with respect to $\overline{h}$ for a fixed but different $h$ for each 3D grid choice.
For large $\overline{h}$, a wire is modeled rather by point sources along $\Lambda$ than by a line source.
For the limit case of $\overline{h}\rightarrow 0$ however, the line source case is recovered and the error with respect to the line source reference solution becomes smaller.
The convergence is of almost second order and independent of the 3D grid and/or grading choice.
This indicates that the 1D discretization error is dominating for the considered error measure and the considered grids.
On the other hand, for the here considered grids, the errors \ensuremath{\varepsilon_{L^{2}}^{\mathrm{1D}}}\xspace, \ensuremath{\varepsilon_{H^{1}}^{\mathrm{1D}}}\xspace and \ensuremath{\delta_{H^{1}}^{\mathrm{1D}}}\xspace are smaller than \num{1e-4} and do not show further improvement for a refinement of the 1D grid.
This behavior is attributed to a dominance of the 3D discretization error for this setting.

\subsubsection{3D-1D Bent Wire Coupling}

We aim for simulations of problems in which thin wires follow arbitrary curves.
Therefore, in this section, we investigate the case of a single bent wire.
Again, we use the parameters of Table~\ref{tab:simSettings} on a cube $D$ as computational domain.
To set up a model of a bent wire with path $\Lambda$ as introduced in Section~\ref{subsec:deRham}, we apply a parametrization using a B\'{e}zier\xspace curve given by
\begin{equation}
    \mathbf{x}(s)=
    \begin{pmatrix}
        d/2\\
        y_{0}(1-s)^{2}+(2y_{0}+4\overline{H})s(1-s)+y_{0}s^{2}\\
        z_{0}(1-s)^{2}+(z_{0}+z_{1})s(1-s)+z_{1}s^{2}
    \end{pmatrix},
    \label{eq:wireBentParam}
\end{equation}
with the start and end coordinates $\mathbf{x}_{0}:=(x_{0},y_{0},z_{0})^{\top}=\mathbf{x}(0)=(0.5,0.02,0.02)^{\top}\si{m}$ and $\mathbf{x}_{1}:=(x_{1},y_{1},z_{1})^{\top}=\mathbf{x}(1)=(0.5,0.02,0.98)^{\top}\si{m}$, respectively, and the bending height $\overline{H}=0.7d$.
The starting and ending points $\mathbf{x}_{0}$ and $\mathbf{x}_{1}$ of the wires are embedded in \gls{PEC} cubes of side length $d_{\text{PEC}}=\SI{40}{mm}$, see Figure~\ref{fig:wireBent}.
This setup is in analogy to the typical case that a wire connects two \gls{PEC} contact pads.

For the construction of the grids, we start with an equidistant 1D grid for the parameter $s$ giving a characteristic 1D step size $\overline{h}$.
From the 1D grid, the 3D wire points are determined by \eqref{eq:wireBentParam} and, together with the boundary nodes in each direction, form the basis of the 3D grid.
Additional grid lines are inserted due to the error evaluation domain $\Omega$, because of the \gls{PEC} cubes and between the wire and the boundary $y=d$.
The number of grid lines used for the latter is given by $\lfloor N^{\mathrm{1D}}/4\rfloor$.

\begin{figure}
    \begin{subfigure}{0.49\columnwidth}
        \includegraphics{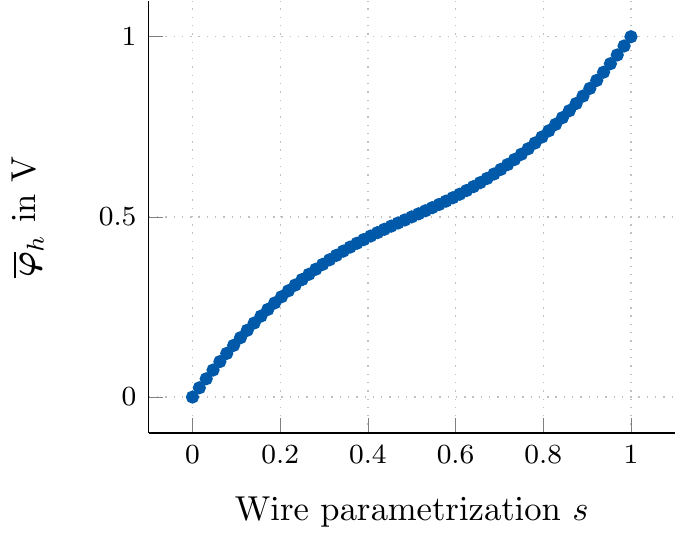}
                \caption{}
        \label{fig:resultsBent1Dplot}
    \end{subfigure}
    \begin{subfigure}{0.49\columnwidth}
        \includegraphics{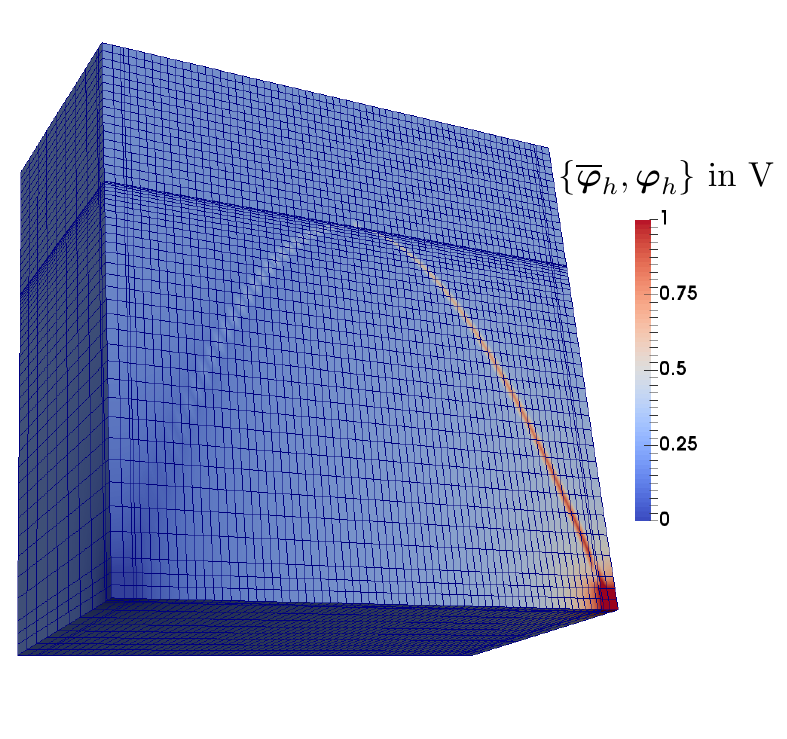}
                \caption{}
        \label{fig:resultsBent3Dplot}
    \end{subfigure}
    \caption{Results for the bent wire 1D-3D coupling for $r_{\mathrm{cpl}}=\num{1e-2}/\overline{\kappa}$. (a) 1D solution $\ensuremath{\boldsymbol{\mathrm{\varphi}}}_{h}$ with respect to the wire parametrization $s$. (b) Values of $\overline{\ensuremath{\boldsymbol{\mathrm{\varphi}}}}_{h}$ on $\Lambda$ and $\ensuremath{\boldsymbol{\mathrm{\varphi}}}_{h}$ on $D\setminus\Lambda$ for the grid sizes of $h\approx\SI{1.884e-2}{m}$ and $\overline{h}\approx\num{1.563e-2}$.}
    \label{fig:resultsBent}
\end{figure}

\begin{figure}
    \begin{subfigure}{0.49\textwidth}
        \includegraphics{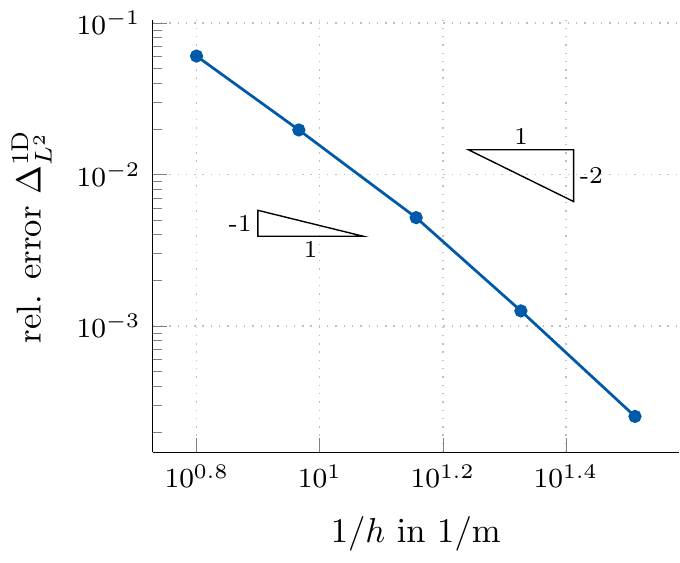}
                \caption{}
        \label{fig:resultsBentDeltaL21D}
    \end{subfigure}
    \begin{subfigure}{0.49\textwidth}
        \includegraphics{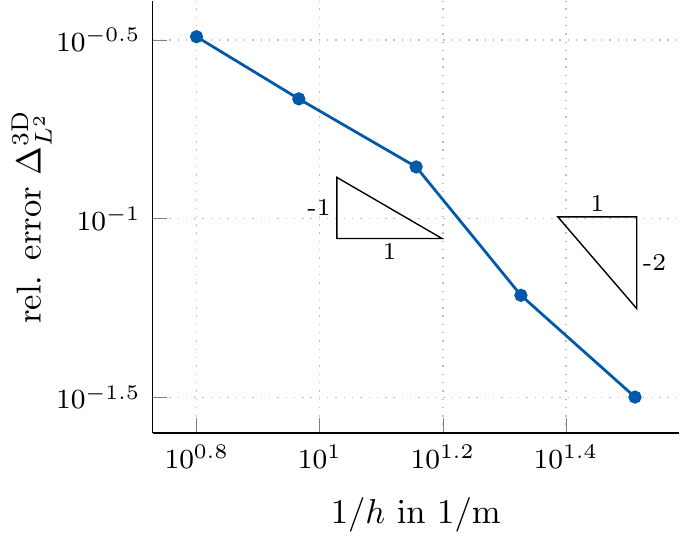}
                \caption{}
        \label{fig:resultsBentDeltaL23D}
    \end{subfigure}
    \caption{Convergence of (a) \ensuremath{\Delta_{L^{2}}^{\mathrm{1D}}}\xspace and (b) \ensuremath{\Delta_{L^{2}}^{\mathrm{3D}}}\xspace with respect to $h$ for the bent wire 1D-3D coupling and ${r_{\mathrm{cpl}}=\num{1e-2}/\overline{\kappa}}$.}
    \label{fig:resultsBentDelta}
\end{figure}

We apply \SI{0}{V} at the \gls{PEC} electrode at $\mathbf{x}_{0}$, \SI{1}{V} at the \gls{PEC} electrode at $\mathbf{x}_{1}$, solve \eqref{eq:electric_deterministic_boundary} and consider the 1D solution as quantity of interest.
We use a coupling radius of $r_{\mathrm{cpl}}=\num{1e-2}/\overline{\kappa}$, where $\overline{\kappa}\approx\SI{6.08}{m^{-1}}$ is the maximum Frenet\xspace-Serret\xspace curvature~\cite{Serret_1851aa,Frenet_1852aa} along the curve.
In the following, we refer to the 1D and 3D reference solutions $\overline{\ensuremath{\boldsymbol{\mathrm{\varphi}}}}$ and $\ensuremath{\boldsymbol{\mathrm{\varphi}}}$ as the solutions computed using $h\approx\SI{1.884e-2}{m}$ and $\overline{h}=\num{1.563e-2}$.
In Figure~\ref{fig:resultsBent1Dplot}, $\overline{\ensuremath{\boldsymbol{\mathrm{\varphi}}}}$ is shown with respect to the wire parametrization $s$ and Figure~\ref{fig:resultsBent3Dplot} shows the solution $\overline{\ensuremath{\boldsymbol{\mathrm{\varphi}}}}_{h}$ on $\Lambda$ and $\ensuremath{\boldsymbol{\mathrm{\varphi}}}_{h}$ on $D\setminus\Lambda$ using a 3D visualization.
Investigating the convergence of the error, we plot \ensuremath{\Delta_{L^{2}}^{\mathrm{1D}}}\xspace and \ensuremath{\Delta_{L^{2}}^{\mathrm{3D}}}\xspace with respect to $h$ in Figure~\ref{fig:resultsBentDelta} using $\overline{\ensuremath{\boldsymbol{\mathrm{\varphi}}}}$ and $\ensuremath{\boldsymbol{\mathrm{\varphi}}}$ as the reference.
For both \ensuremath{\Delta_{L^{2}}^{\mathrm{1D}}}\xspace and \ensuremath{\Delta_{L^{2}}^{\mathrm{3D}}}\xspace, we observe a convergence order of around two.

\subsection{Industry-Relevant Microelectronic Chip Package Simulation}

We consider a microelectronic chip package model based on~\cite{Casper_2016aa}, see Figure~\ref{fig:chip_package}.
The contact pads are centered in $z$-direction with a height of \SI{100}{\mu m}.
The radius of all wires is $\overline{r}=\SI{1}{\mu m}$ with a circular cross-sectional area given by $\overline{A}=\pi\overline{r}^2$.
The curvature of the wires is given by \eqref{eq:wireBentParam} and a height of $\overline{H}=\SI{0.1}{mm}$, where $\mathbf{x}_{0}$ and $\mathbf{x}_{1}$ differ for the different wires. 

\begin{table}[p]
    \centering
    \begin{tabular}{ccc}\toprule
        Parameter               & Description                                    & Value\\\bottomrule\toprule
        $\overline{r}$                & Radius of all wires                            & \SI{1}{\mu m}\\
        $\overline{A}$                & Cross-sectional area of wire                   & $\pi\overline{r}^{2}$\\
        $\overline{H}$                & Height of all wires                            & \SI{0.1}{mm}\\
        $\sigma_{\mathrm{con}}$             & Electric conductivity of $D_{\mathrm{con}}$      & $\infty$ (PEC)\\
        $\sigma_{\mathrm{ins}}$             & Electric conductivity of $D_{\mathrm{ins}}$      & \SI{1e-4}{S\per\metre}\\
        $\lambda_{\mathrm{con}}$            & Thermal conductivity of $D_{\mathrm{con}}$       & $\lambda_{\text{Cu}}$\\
        $\lambda_{\mathrm{ins}}$            & Thermal conductivity of $D_{\mathrm{ins}}$       & \SI{0.87}{W\per(Km)}\\
        $\rho_{\mathrm{con}}$           & Volumetric mass density of $D_{\mathrm{con}}$    & $\rho_{\text{Cu}}$\\
        $\rho_{\mathrm{ins}}$           & Volumetric mass density of $D_{\mathrm{ins}}$    & \SI{1500}{kg\per m\tothe{3}}\\
        $c_{\mathrm{con}}$              & Specific heat density of $D_{\mathrm{con}}$      & $c_{\text{Cu}}$\\
        $c_{\mathrm{ins}}$              & Specific heat density of $D_{\mathrm{ins}}$      & \SI{882}{J\per(Kkg)}\\
        $\overline{\sigma}$            & Electric conductivity of $\Lambda$             & $|\overline{A}|\sigma_{\text{Cu}}$\\
        $\overline{\lambda}$           & Thermal conductivity of $\Lambda$              & $|\overline{A}|\lambda_{\text{Cu}}$\\
        $\varphi_{\text{init}}$ & Initial potential                              & \SI{0}{V}\\
        $T_{\text{init}}$       & Initial temperature                            & \SI{300}{K}\\
        $h$                     & Heat transfer coefficient                      & \SI{25}{W\per(\metre\squared K)}\\
        $T_{\infty}$            & Ambient temperature                            & \SI{300}{K}\\
        $\overline{V}$                & Wire voltages                                  & \SI{100}{mV}\\
        $r_{\mathrm{cpl}}$                 & Coupling radius                                & $\num{1e-4}\overline{H}^{2}\overline{\kappa}$\\
        $N^{\mathrm{1D}}$                & No.\ of 1D points                              & \num{4}\\
        $N_{t}$                 & Number of time points                          & \num{10}\\
        $t_{0}$                 & End time                                       & \SI{1}{s}\\
        \bottomrule
    \end{tabular}
    \caption{Parameters used for the chip package simulation.}
    \label{tab:chipParameters}
\end{table}

\begin{figure}
    \centering
    \begin{subfigure}{0.55\textwidth}
        \includegraphics{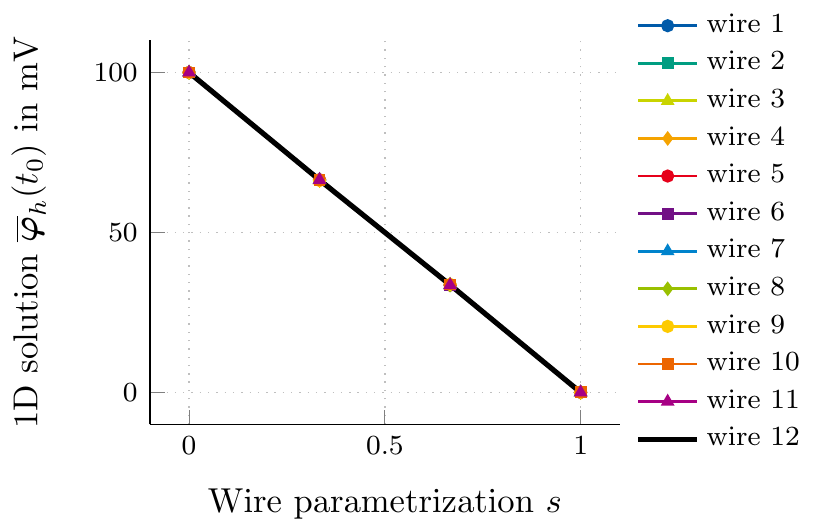}
                \caption{}
        \label{fig:resultsChipElectric1D}
    \end{subfigure}
    \begin{subfigure}{0.44\textwidth}
        \includegraphics{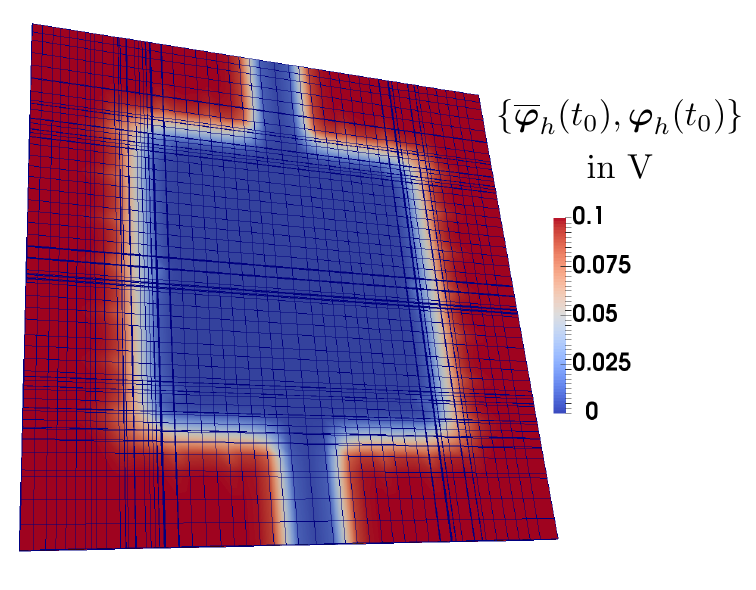}
                \caption{}
        \label{fig:resultsChipElectric3D}
    \end{subfigure}
    \caption{Resulting electric solution for the electrothermal chip package simulation using grids with ${h\approx\SI{8.51e-5}{m}}$ and $\overline{h}=\num{1/3}$. (a) shows the 1D solution $\overline{\ensuremath{\boldsymbol{\mathrm{\varphi}}}}_{h}(t_{0})$ for the \num{12} wires attached to the 3D chip package (for wire numbering see Figure~\ref{fig:chip_package}). (b) shows the 3D solution $\ensuremath{\boldsymbol{\mathrm{\varphi}}}_{h}(t_{0})$ at $z=\SI{150}{\mu m}$ in the $xy$-plane.}
    \label{fig:resultsChipElectric}
\end{figure}

\begin{figure}
    \centering
    \begin{subfigure}{0.55\textwidth}
        \includegraphics{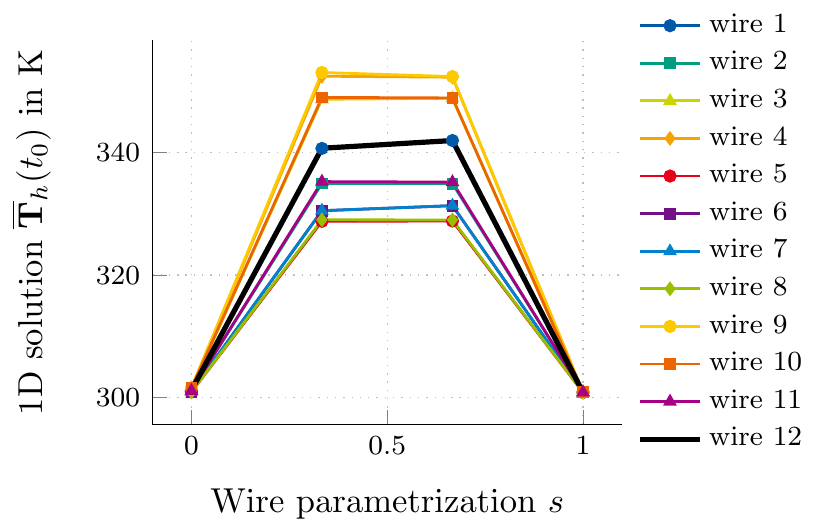}
                \caption{}
        \label{fig:resultsChipThermal1D}
    \end{subfigure}
    \begin{subfigure}{0.44\textwidth}
        \includegraphics{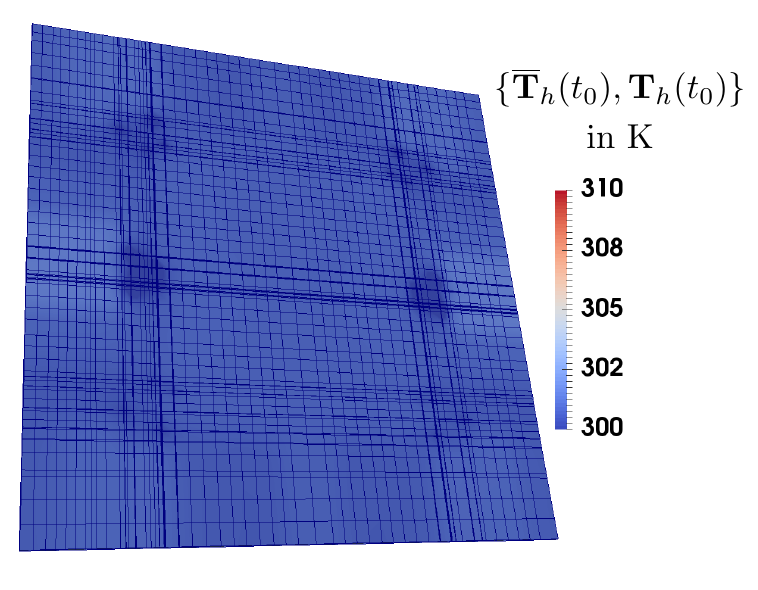}
                \caption{}
        \label{fig:resultsChipThermal3D}
    \end{subfigure}
    \caption{Resulting thermal solution for the electrothermal chip package simulation using grids with ${h\approx\SI{8.51e-5}{m}}$ and $\overline{h}=\num{1/3}$. (a) shows the 1D solution $\overline{\ensuremath{\mathbf{T}}}_{h}(t_{0})$ for the \num{12} wires attached to the 3D chip package (wire numbering cf. Figure~\ref{fig:chip_package}). (b) shows the 3D solution $\ensuremath{\mathbf{T}}_{h}(t_{0})$ at $z=\SI{150}{\mu m}$ in the $xy$-plane.}
    \label{fig:resultsChipThermal}
\end{figure}

For simplicity, we assume all material characteristics to be linear and model $D_{\mathrm{con}}$ with an electric conductivity $\sigma_{\mathrm{con}}\to\infty$ (\gls{PEC}) and $D_{\mathrm{ins}}$ with $\sigma_{\mathrm{ins}}=\SI{1e-4}{S\per\metre}$.
On the other hand, the thermal conductivities are given by $\lambda_{\mathrm{con}}=\lambda_{\text{Cu}}$ and $\lambda_{\mathrm{ins}}=\SI{0.87}{W\per(K\metre)}$, where $\lambda_{\text{Cu}}=\SI{401}{W\per(K\metre)}$ is the conductivity of copper.
The volumetric mass densities and specific heat capacities are given by $\rho_{\mathrm{con}}=\rho_{\text{Cu}}$, $\rho_{\mathrm{ins}}=\SI{1500}{kg\per m\tothe{3}}$, $c_{\mathrm{con}}=c_{\text{Cu}}$ and $c_{\mathrm{ins}}=\SI{882}{J\per(Kkg)}$, respectively, where $\rho_{\text{Cu}}=\SI{8930}{kg\per m\tothe{3}}$ and $c_{\text{Cu}}=\SI{390}{J\per(Kkg)}$ are the values for copper.
We choose the wires' conductivities to be $\overline{\sigma}=\overline{A}\sigma_{\text{Cu}}$ and $\overline{\lambda}=\overline{A}\lambda_{\text{Cu}}$, where $\sigma_{\text{Cu}}=\SI{5.96e7}{S\per\metre}$ is the conductivity of copper.
As initial conditions, the package is homogeneously set to $\varphi_{\text{init}}=\SI{0}{V}$ and $T_{\text{init}}=\SI{300}{K}$.
A voltage of $\overline{V}=\SI{100}{mV}$ is applied over each wire, where the central chip region is used as the ground potential.
In terms of electrical boundary conditions, this translates to Dirichlet\xspace conditions on $\Gamma_{\mathrm{Dir},l}$ and homogeneous Neumann\xspace conditions on $\partial D\setminus\Gamma_{\mathrm{Dir},l}$.
Thermal boundary conditions are chosen to be convective on $\partial D$ with the heat transfer coefficient $h=\SI{25}{W\per(\metre\squared K)}$ and the ambient temperature $T_{\infty}=\SI{300}{K}$.
For each of the wires, the 1D-3D coupling is carried out using $r_{\mathrm{cpl}}=\num{1e-4}\overline{H}^{2}\overline{\kappa}$ and $N^{\mathrm{1D}}=4$.
All mentioned simulation parameters are summarized in Table~\ref{tab:chipParameters}.
The applied grid results from a commercial meshing tool for the 3D part and a subsequent insertion of grid lines at the points given by~\eqref{eq:wireBentParam} to yield $h\approx\SI{8.51e-5}{m}$.
The time discretization is given by $N_{t}=\num{10}$ with $t_{0}=\SI{1}{s}$.
For a choice of $\overline{h}=\num{1/3}$, we solve~\eqref{eq:electrothermal_deterministic_boundary} and obtain the results for $\ensuremath{\boldsymbol{\mathrm{\varphi}}}_{h}(t_{0})$, $\overline{\ensuremath{\boldsymbol{\mathrm{\varphi}}}}_{h}(t_{0})$, $\ensuremath{\mathbf{T}}_{h}(t_{0})$ and $\overline{\ensuremath{\mathbf{T}}}_{h}(t_{0})$ as shown in Figure~\ref{fig:resultsChipElectric} and \ref{fig:resultsChipThermal}.
The presented results demonstrate the capability to predict the temperature distribution in a microelectronic chip package including the temperature profile of the bond wires.

\section{Conclusion}

Motivated by the electrothermal simulation of microelectronic chip packages including thin bond wires, we presented an approach that alleviates the necessity of fine grids to resolve the geometry of the wires.
In the literature, such a problem is known as a 1D-3D coupling and was, e.g., investigated by~\cite{DAngelo_2008aa} in the context of fluid flow in porous media with fractures.
The main challenge of this problem was identified to be the solution-dependent line source term in the 3D Laplace\xspace equation.
Using the setting of de Rham\xspace currents to describe the arising singularities, we proposed a continuous formulation of the 1D-3D electrothermal coupling.
The coupling condition follows the work in~\cite{DAngelo_2008aa} to average the 3D solution around a wire for the definition of the 1D solution.
Additionally, we introduced a scaling factor that accounts for the distance from the wire to yield a physical solution for small but arbitrary coupling radii.

The discrete system was set up using the discrete counterpart of de Rham\xspace currents in a \gls{FIT} formulation.
A detailed description of all the involved discretization steps was presented.
The theory concludes by showing that the relation of the present thin wire problem to \gls{FEM} problems for fluid flow problems with fractures is established by an infinite permeability of the vessel-tissue interface.

To investigate numerical errors and convergence rates, electric model problems for a 0D-2D coupling and a straight and bent wire 1D-3D coupling were considered.
The convergence rates of the corresponding errors were analyzed for different choices of the grid, the grid grading and the coupling radius.
It was shown that the grid grading can improve the convergence rate substantially while the solution is independent of the coupling radius.
We could see in particular that, due to the singularity of the 3D solution at the 1D domain, a direct coupling of 1D- and 3D-points ($r_{\mathrm{cpl}}=0$) results in a very slow convergence.
Lastly, a transient electrothermal simulation of a chip package with 12 applied wires was presented.

The numerical analysis of the nonlinear, transient and coupled electrothermal problem is still an open research topic.
In particular, existence and uniqueness need to be established for case of an infinite permeability, which is not included in the analysis presented in~\cite{DAngelo_2012aa}.
Additionally, the error analysis for the special case of the \gls{FIT} discretization is of interest.

\section*{Acknowledgment}
The authors thank Winnifried Wollner for the fruitful discussions on the topic.
This work is supported by the European Union within FP7-ICT-2013 in the context of the \emph{Nano-electronic COupled Problems Solutions} (nanoCOPS) project (grant no. 619166), by the \emph{Excellence Initiative} of the German Federal and State Governments and the Graduate School of CE at Technische Universität Darmstadt.

\appendix

\end{document}